\documentclass{LMCS}

\def\dOi{10(1:5)2014}
\lmcsheading%
{\dOi}
{1--26}
{}
{}
{Aug.~\phantom07, 2013}
{Feb.~\phantom07, 2014}
{}

\ACMCCS{[{\bf Theory of computation}]:  Logic---Logic and
  verification\,/\,Equational logic and rewriting}
\usepackage{epsfig, amssymb, amsxtra}
\usepackage{latexsym, graphics, graphicx}
\usepackage{algorithmic, algorithm} 
\usepackage{proof}
\usepackage{hyperref}
\newcommand{\ignore}[1]{}

\newcommand{\flr}{\rightarrow}

\newcommand{\lft}{\noindent}
\newcommand{\disp}[1] { \vspace*{0.3em}
      \begin{center} { #1 }  \end{center}  \vspace*{0.3em} }

\newcommand{\A}{\mathcal{A}}

\newcommand{\E}{\mathcal{E}}
\newcommand{\T}{\mathcal{T}}
\newcommand{\X}{\mathcal{X}}

\newcommand{\p}{\mathcal{P}}

\newcommand{\eq}{\mathcal{EQ}}
\newcommand{\Eq}{=_{}^?}
\newcommand{\bc}{\mathcal{BC}}
\newcommand{\dbc}{\mathcal{DBC}}
\newcommand{\Inf}{\mathcal{INF}}

\newcommand{\hsp}{\hspace*{2em}}

\newcommand{\hspa}{\hspace*{1cm}}

\begin{document}

\title[Unification modulo a 2-sorted Equational theory for
  CBC]{Unification modulo a 2-sorted Equational theory \\ for
  Cipher-Decipher Block Chaining}

\author[S.~Anantharaman]{Siva Anantharaman\rsuper a}
\address{{\lsuper a} LIFO, Universit\'e d'Orl\'eans (France)}
\email{siva@univ-orleans.fr}

\author[C.~Bouchard]{Christopher Bouchard\rsuper b}
\address{{\lsuper{b,c}}University at Albany--SUNY (USA)}
\email{\{cb829983,dran\}@cs.albany.edu }
\thanks{{\lsuper{b,c}}Research supported in part by NSF grant CNS-0905286} 

\author{Paliath  Narendran\rsuper c}
\address{\vspace{-18 pt}}

\author[M.~Rusinowitch]{Micha\"{e}l Rusinowitch\rsuper d}
\address{{\lsuper d}Loria-INRIA Grand Est, Nancy (France)}
\email{rusi@loria.fr}
\thanks{{\lsuper d} Research supported in part by FP7 NESSOS Project} 

\begin{abstract}
We investigate unification problems related to the Cipher Block Chaining 
(CBC) mode of encryption.
We first model chaining in terms of a simple, convergent, rewrite system 
over a signature with two disjoint sorts: {\em list\/} and {\em element.\/} 
By interpreting a particular symbol of this signature suitably, the rewrite 
system can model several practical situations of interest. 
An inference procedure is presented for deciding the unification problem 
modulo this rewrite system. The procedure is modular in the following sense: 
any given problem is handled by a system of `list-inferences', and the set 
of equations thus derived between the element-terms of the problem is then 
handed over to any (`black-box') procedure which is complete for solving 
these element-equations.  
An example of application of this unification procedure is given, as 
attack detection on a Needham-Schroeder like protocol, employing the CBC 
encryption mode based on the associative-commutative (AC) operator XOR.  
The 2-sorted convergent rewrite system is then extended into one that fully 
captures a block chaining encryption-decryption mode at an abstract level, 
using no AC-symbols; and unification modulo this extended system is 
also shown to be decidable. 
\end{abstract} 

\keywords{Equational unification, Block chaining, Protocol}

\maketitle

\section{Introduction}~\label{intro}

The technique of {\em chaining\/} is applicable in many situations. A simple 
case is e.g., when we want to calculate  the partial sums (resp.\ products) 
of a (not necessarily bounded) list of integers, with a given `base' integer; 
such a list of partial sums (resp. products) can be calculated, incrementally, 
with the help of  the following two equations:  
\disp{$bc(nil, \, z) = nil, \; \hsp  
           bc(cons(x, Y), \, z)  =  cons(h(x,z), \; bc(Y, \, h(x,z)))$}
where $nil$ is the empty list, $z$ is the given base integer, $x$ is an 
integer variable, and $Y$ is the given list of integers. The partial 
sums (resp. products) are returned as a list, by evaluating the function 
$bc$, when $h(x, z)$ is interpreted as the sum (resp.\ product) of $x$ with 
the given base integer $z$. 

A more sophisticated example is the Cipher Block Chaining encryption mode 
(CBC, in short), employed in cryptography, a mode which uses the AC-operator 
exclusive-or (XOR) for `chaining the ciphers across the message blocks'; here 
is how this is done: Let~$\oplus$ stand for XOR (which we let distribute over
block concatenation), and let $M = p_1  \dots p_n$ be a message given as 
a list of $n$ `plaintext' message subblocks. Then the encryption of $M$, 
with any given public key $k$ and an initialization vector $v$, is defined 
as the list $c_1 \dots c_n$ of ciphertext message subblocks, where:
$c_1 = e_k(p_1 \oplus v)$, \, and ~~$c_i = e_k(p_i \oplus c_{i-1})$,
     for any  $1 < i \le n$. 
(Note: It is usual in Cryptography to see a message as a sequence of 
``records'', each record being decomposed into a sequence of blocks of 
the same size; what we refer to as `message' in this paper, 
would then correspond to a `record' in the sense of cryptography.) 
The above set of equations also models this  CBC encryption mode: for this,
we interpret the function $h(x, y)$ as the encryption $e_k(x \oplus y)$ of 
any single block message $x$,  XOR-ed with the initialization vector $y$, 
using the given public key $k$. Under such a vision, a message  $M$ is 
decomposed as the concatenation of its first message block $m$ with the 
rest of the message list $M'$, i.e., we write $M = m \cdot M'$; then,  
the encryption of $M$ with any given public key $k$, with $x$ taken as 
initialization vector (IV), is derived by 
$bc(M, x) = h(m, x) \cdot bc(M', h(m, x))$. 

Actually, our interest in the equational theory defined by the above 
two equations was motivated by the possibility of such a modeling for 
Cipher Block Chaining, and the fact that rewrite as well as unification 
techniques are often employable, with success, for the formal analysis 
of cryptographic protocols (cf. e.g.,
\cite{AbadiCortier2004,siva-dran-micha-07,Baudet05,Comon-T03,Comon-LICS03},
and also the concluding section).   

This paper is organized as follows. In Section~\ref{prelim} we introduce 
our notation and the basic notions used in the sequel; we shall observe,  
in particular, that the two equations above can be turned into rewrite 
rules and form a convergent rewrite system over a 2-sorted signature: 
{\em lists\/} and {\em elements}. 
Our concern in Section~\ref{bc-inf} is the unification problem modulo 
this rewrite system, that we denote by $\bc$; we present a 2-level inference 
system (corresponding, in a way, to the two sorts of the signature) for 
solving this problem. 
Although our main aim is to investigate the unification problem for the 
case where $h$ is an interpreted function symbol (as in the two situations 
illustrated above), we shall also be considering the case where $h$ 
is a free uninterpreted symbol. The soundness and completeness of our 
inference procedure are established in Section~\ref{method}. 
While the complexity of the unification problem is polynomial over the 
size of the problem when $h$ is uninterpreted, it turns out to be 
NP-complete when $h$ is interpreted so that the rewrite system models 
 CBC encryption. We then present, in Section~\ref{DBC}, 
a 2-sorted convergent system $\dbc$ that fully models at an abstract 
level, a block chaining cipher-decipher mode without using any 
AC-operators; this is done by adding a couple of equations to the above 
two: one for specifying a left-inverse $g$ for $h$  ($g$ does the 
deciphering), and the other for specifying the block chaining mode for 
deciphering. A 2-level inference procedure extending the one given in 
Section~\ref{bc-inf} is presented, and is shown to be sound and complete 
for unification modulo this extended system $\dbc$; unification modulo $\dbc$ 
also turns out to be NP-complete. In the concluding section we briefly evoke 
possible lines of future work over these systems $\bc$ and $\dbc$. 

\medskip
{\em Note:} The first part of this paper, devoted to unification modulo 
$\bc$, is a more detailed version of the work we presented at 
LATA 2012~(\cite{cbcreport}).  

\section{Notation and Preliminaries}\label{prelim}

We consider a ranked signature $\Sigma$, with two {\em disjoint\/} sorts:   
$\tau_e^{}$ and $\tau_l^{}$, consisting of binary functions 
{\em bc, cons, h\/}, and a constant $nil$, and typed as follows: 
\disp{ 
   $bc : \; \tau_l^{}  \times \tau_e^{} \rightarrow \tau_l^{}$ \, ,
  \, $cons : \; \tau_e^{}  \times \tau_l^{} \rightarrow \tau_l^{}$\, ,  
  \, $h : \; \tau_e^{}  \times \tau_e^{} \rightarrow \tau_e^{}$\, , 
  \, $nil: \tau_l^{}$.}
We also assume given a set $\X$ of countably many variables; the 
objects of our study are the (well-typed) terms of the algebra 
$\T(\Sigma, \X)$; terms of the type $\tau_e^{}$ will be referred to 
as {\em elements\/}; and those of the type $\tau_l^{}$  
as {\em lists}. It is assumed that the only constant of type list is $nil$;  
the other constants, if any, will all be of the type element. 
For better readability, the set of variables $\X$ will be divided 
into two subsets: those to which `lists' can get assigned  
will be denoted with upper-case letters as: $X, Y, Z, U, V, W, \dots$, 
with possible suffixes or primes; these will be said to be variables 
of type $\tau_l^{}$; variables to which `elements' can get assigned  will 
be denoted with lower-case letters, as: $x, y, z, u, v, w, \dots$, with 
possible suffixes or primes; these will be said to be variables of type 
$\tau_e^{}$.  The theory we shall be studying first in this paper is defined 
by the two axioms (equations) already mentioned in the Introduction:  

\disp{$bc(nil, \, z) = nil, \; \hsp  
         bc(cons(x, Y), \, z)  =  cons(\, h(x,z), \; bc(Y, \, h(x,z)) \,)$}

It is easy to see that these  axioms can both be oriented left-to-right 
under a suitable {\em lexicographic path ordering (lpo)\/}~(cf. 
e.g.,~\cite{Dersh-JSC}), and that they form then a convergent --- i.e., 
confluent and terminating --- 2-sorted rewrite system.  

As mentioned in the previous section, we consider two theories
that contain the above two axioms. The first is where these are
the {\em \/} only axioms; we call that theory~${\bc}_0^{}$. The other
theory is where $h$~is interpreted as for CBC, i.e., where 
$h(x, y) = e_k^{} (x \oplus y)$ where $\oplus$~is exclusive-or and $e_k^{}$
is encryption using some  (fixed) given key~$k$. This theory will be 
referred to as~${\bc}_1^{}$. We use the phrases ``$\bc$-unification''
and ``unification modulo $\bc$'' to refer to unification problems
modulo both the theories, collectively.

Note that in the case where $h$ is a free uninterpreted symbol
(i.e.,~${\bc}_0^{}$) $h$ is fully cancellative in the sense that
for any terms $s_1, t_1, s_2, t_2$, $h(s_1, t_1) ~
{\approx}_{\bc}^{} ~ h(s_2, t_2)$
if and only if $s_1 ~ {\approx}_{\bc}^{} ~ s_2$ and $t_1 ~
{\approx}_{\bc}^{} ~ t_2$.  
But when $h$ is interpreted for CBC, this is no longer true;  in such 
a case, $h$ will be only {\em  semi-cancellative\/}, in the sense that 
for all terms $s_1, s_2, t$, the following holds:

$h$ is right-cancellative: $h(s_1, t) ~ {\approx}_{\bc}^{} ~ h(s_2, t)$
   if and only if $s_1 ~ {\approx}_{\bc}^{} ~ s_2$,  and \par 
$h$ is also left-cancellative: $h(t, s_1) ~ {\approx}_{\bc}^{} ~ h(t, s_2)$
   if and only if $s_1 ~ {\approx}_{\bc}^{} ~ s_2$. 

Thus, in the sequel, when we look for the unifiability of any set of element 
equations modulo $\bc_0$ (resp. modulo $\bc_1$) the cancellativity of $h$ 
(resp. the semi-cancellativity of $h$) will be used as needed, in general 
without any explicit mention.  

\medskip{}
Our concern in this section, and the one following, is the equational 
unification problems modulo ${\bc}_0^{}$ and ${\bc}_1^{}$. We assume 
without loss of generality (wlog) that any given $\bc$-unification 
problem $\p$ is in {\em standard form,\/} i.e., $\p$ is given as a 
set of equations $\eq$, each having one of the following forms: 

\disp{$U =_{}^? V, \; U =_{}^? bc(V, y),\; U =_{}^? cons(v, W), 
                     \; U =_{}^? nil, \;$ \\
         $u =_{}^? v, \; \; v=_{}^? h(w,x), \; u \Eq const$}
where $const$ stands for any ground constant of sort~$\tau_e^{}$. The 
first four kinds of equations --- the ones with a list-variable 
on the left-hand side --- are called {\em list-equations,\/}
and the rest (those which have an element-variable on the left-hand side) 
are called {\em element-equations.\/}
For any problem $\p$ in standard form, $\mathcal{L}(\p)$ will 
denote the  subset formed of its list-equations, and $\mathcal{E}(\p)$  
the subset of element-equations. A set of element-equations is said to be
in {\em dag-solved form\/} (or {\em d-solved form\/}) (\cite{JoKi}) if and 
only if they can be arranged as a list $x_1 =_{}^? t_1 , \; \ldots , \; 
 x_n =_{}^? t_n$, such that: \par 
\disp{ $\forall \, 1 \le i < j \le n$: ~~ $x_i$ and $x_j$ are 
distinct variables, and $x_i$ does not occur in $t_i$ nor in any $t_j$. }  
Such a notion is naturally extended to sets of list-equations as well.   
In the next section we give an inference system for solving any
$\bc$-unification problem in standard form. 
For any given problem $\p$, its rules will transform 
$\mathcal{L}(\p)$ into one in $d$-solved form. The element-equations at 
that point can be passed on to an algorithm for solving them --- 
thus in the case of~${\bc}_1^{}$ what we need is an algorithm for solving
the {\em general\/} unification problem modulo the theory of exclusive-or.

Any development presented below --- without further precision on
$h$ --- is meant as one which will be valid for both $\bc_0^{}$
and $\bc_1^{}$.

\section{Inference System for $\bc$-Unification}~\label{bc-inf}

The inference rules have to consider two kinds of equations: 
the rules for the {\em list-equations\/} in $\p$, i.e., equations 
whose left-hand sides (lhs) are variables of type $\tau_l^{}$, and 
the rules for the {\em element-equations,\/} i.e., equations whose 
lhs are variables of type $\tau_e^{}$. 
Our method of solving any given unification problem will be `modular'  
on these two sets of equations: The list-inference rules will be shown to 
terminate under suitable conditions, and then all we will need to do is 
to solve the resulting set of element-equations for $h$.  

A few technical points need to be mentioned before we formulate our 
inference rules. Note first that it is not hard to see that $cons$ is 
cancellative; by this we mean that  
$cons(s_1, T_1) ~ {\approx}_{\bc}^{} ~ cons(s_2, T_2)$, 
for terms $s_1^{}, \, s_2^{}, \, T_1^{}, \, T_2^{}$, if and only if 
$s_1 ~ {\approx}_{\bc}^{} ~ s_2$ and $T_1 ~ {\approx}_{\bc}^{} ~ T_2$. 
On the other hand, it can be shown by structural induction (and the 
semi-cancellativity of $h$) that $bc$ is {\em conditionally\/} 
semi-cancellative, depending on whether its first argument is $nil$ or not; 
for details, see  {\em Appendix-1}. This property of $bc$ will be 
assumed in the sequel.  

\medskip
The inference rules given below will have to account for cases where an  
`occur-check' succeeds on some list-variable, and the problem will be 
unsolvable. The simplest among such cases is when we have an equation of 
the form $U =_{}^? cons(z,U)$ in the problem. But one could have more complex 
unsolvable cases, where the equations involve  both $cons$ and $bc$; e.g., 
when $\p$ contains equations of the form: 
$U =_{}^? cons(x,V), U =_{}^? bc(V, y)$; the problem will be unsolvable in 
such a case: indeed, from the axioms of $\bc$, one deduces that $V$ must 
be of the form $V =_{}^? cons(v, V')$, for some $v$ and $V'$, then  $x$ 
must be of the form $x =_{}^? h(v, y)$, and subsequently $V =_{}^? bc(V', x)$, 
and we are back to a set of equations of the same format. 
We need to infer failure in all such cases. With that purpose, we define the 
following relations on the list-variables of the equations in $\p$: 

\begin{itemize}
  \item $U >_{cons} V$ iff $U =_{}^? cons(z,V)$, for some $z$.
  \item $U ~>_{bc} ~V$ iff there is an equation  $U =_{}^? bc(V, x)$
  \item $U \sim_{bc} V$ iff $U =_{}^? bc(V,w)$, or $V =_{}^? bc(U,w)$, 
       for some $w$.  
\end{itemize}
Note that $\sim_{bc}^{}$ is the symmetric closure of the relation $>_{bc}$;
its reflexive, symmetric and transitive closure is denoted as
$\sim_{bc}^{*}$. The transitive closure of $>_{bc}$ is denoted as $>_{bc}^{+}$; 
and its reflexive transitive closure as $>_{bc}^{*}$.

Note, on the other hand, that $U \Eq bc(U, x)$ is solvable by the 
substitution $\{ U := nil \}$; in fact this equation forces $U$ to be $nil$, 
as would also a set of equations of the form  
$U =_{}^? bc(V, y), ~ V =_{}^? bc(U, x)$. Such cycles (as well as some others) 
have to be checked to determine whether a list-variable  is forced to be 
$nil$. This can be effectively done with the help of the relations 
defined above on the type $\tau_l^{}$ variables. We define,  recursively, 
a set {\bf nonnil} of the list-variables of $\p$ that cannot be  $nil$ for 
any unifying substitution, as follows: 

\begin{itemize}
\item if $U =_{}^? cons(x, V)$ is an equation in $\p$, then 
$U \in \mathbf{nonnil}$.

\item if $U =_{}^? bc(V, x)$ is an equation in $\p$, then 
    $U \in \mathbf{nonnil}$ if and only if $V \in \mathbf{nonnil}$.
\end{itemize}

\lft
We have then the following obvious result: 
\begin{lem} \label{nonnilvar}
A variable $U \in \mathbf{nonnil}$ if and only if there are variables 
$V$ and~$W$ such that $U \sim_{bc}^* V$ and $V >_{cons}^{} W$.  
\end{lem}

Some of the inference rules below will refer to a graph whose nodes are 
the list-variables of the given problem $\p$, `considered equivalent   
up to equality'; more formally:  for any list-variable $U$ of $\p$, we denote 
by $[U]$ the equivalence class of list-variables that get equated to $U$ 
in $\p$, in the following sense: 
\disp{$[U] = \{ V \mid U =_{}^? V \in \p~\mathrm{or}~V =_{}^? U \in \p\}.$}
Any relation $\mathcal{R}$ defined over the list-variables of $\p$ is then 
extended naturally to these equi\-valence classes, by setting:
$\mathcal{R}([U_1], \dotsc, [U_n]) \; ~ \mathrm{iff}~ \; 
  \exists V_1 \in [U_1] \, \dotso \, \exists V_n \in [U_n] 
  \colon \mathcal{R}(V_1, \dotsc, V_n)$.

\begin{defi}
Let $G_l = G_l(\p)$ be the graph whose nodes are the equivalence classes on
the list-variables of $\p$, with arcs defined as follows: From a node 
$[U]$ on $G_l$ there is a {\em directed\/} arc to a (not necessarily 
different) node $[V]$ on $G_l$ if and only if:
\begin{itemize}
\item Either $U \, >_{cons} \, V$: in which case the arc is labeled with $>_{cons}$
\item $U  >_{bc}  V$: in which case the arc is labeled with  $>_{bc}$.
\end{itemize}
\noindent
In the latter case, $G_l$ will also have a {\em two-sided (undirected)\/} 
edge between $[U]$ and $[V]$, which is labeled with  $\sim_{bc}$.
The graph $G_l$ is called the {\em propagation graph\/} for $\p$. 
\end{defi}
  
A node $[U]$ on $G_l$ is said to be a $bc/bc$-peak if $\p$ contains two 
different equations of the form  $U \Eq bc(V, x), U \Eq bc(W, y)$; the node 
 $[U]$ is said to be a $cons/bc$-peak if $\p$ has two different equations 
of the form  $U \Eq cons(x, V_1), \, U \Eq bc(V, z)$.

\medskip
On the set of nodes of $G_l$, we define a partial relation $\succ_l$ by 
setting: $[U] \succ_l [V]$ iff there is a path on $G_l$ from 
$[U]$ to $[V]$, at least one arc of which has label $>_{cons}$.
In other words,
\disp{$\succ_l ~ = ~ \sim_{bc}^* \; \circ \; >_{cons} \; 
  \circ \; (\sim_{bc} \cup >_{cons})_{}^{*}$}
A list-variable $U$ of $\p$ is said to \emph{violate occur-check\/} iff
$[U] \succ_l [U]$ on $G_l$. 
For instance, the variable $U$ violates occur-check in the problem: 
\disp{$U \Eq bc(W, z), \, W \Eq cons(x, U)$,} 
as well as in the problem: 
\disp{$ U \Eq bc(V, z), \,  V \Eq bc(W, a), W \Eq cons(a, L), 
         \, L \Eq bc(U, b)$}
It can be checked that both the problems are unsatisfiable. 

\subsection{Inference System $\Inf_l^{}$ for List-Equations}~\label{Inf-l}

\medskip{}
\lft (L1) {\em Variable Elimination}:\par
\centerline{
$\infer[\qquad \mathrm{if} ~ U ~ \mathrm{occurs ~ in} ~
           \eq ]{\{U =_{}^? V\} \cup \, [V/U](\eq) }
               { \{U =_{}^? V\} ~ \uplus ~ \eq }$
}

\vspace*{0.3em}
\lft (L2) {\em Cancellation on $cons$}:\par
\centerline{
$\infer{\eq ~ \cup ~ \{ U =_{}^? cons(x, V), \; v =_{}^? x, \;
                                                W =_{}^? V \}}
  {\eq ~ \uplus ~ \{ U =_{}^? cons(v, W), \; U =_{}^? cons(x, V) \}}$
}

\vspace*{0.3em}
\lft (L3.a) {\em Nil solution-1}:\par
\centerline{
$\infer{\eq ~ \cup ~ \{ U =_{}^? nil, \; V =_{}^? nil \}}
  {\eq ~ \uplus ~ \{ U =_{}^? bc(V, x), \; U =_{}^? nil \}}$
}

\vspace*{0.3em}
\lft (L3.b) {\em Nil solution-2}:\par
\centerline{
$\infer{\eq ~ \cup ~ \{ U =_{}^? nil, \; V =_{}^? nil \}}
 {\eq ~ \uplus ~ \{ U =_{}^? bc(V, x), \; V =_{}^? nil \}}$
}

\vspace*{0.3em}
\lft (L3.c) {\em Nil solution-3}:\par
\centerline{
$\infer[\qquad \mathrm{if} ~ V >_{bc}^{*} U ]
   {\eq ~ \cup ~ \{ U =_{}^? nil, \; V =_{}^? nil \}}
   {\eq ~ \uplus ~ \{ U =_{}^? bc(V, x) \}}$
}

\vspace*{0.3em}
\lft (L4.a) {\em Semi-Cancellation on $bc$}, at a $bc/bc$-peak:\par
\centerline{
$\infer
   {\eq ~ \cup ~ \{U =_{}^? bc(V, x), \; W =_{}^? V \}}
   {\eq ~ \uplus ~ \{ U =_{}^? bc(V, x), \; U =_{}^? bc(W, x) \}}$
}

\vspace*{0.3em}
\lft (L4.b) {\em Push $bc$ below $cons$}, at a 
                        $\mathbf{nonnil}$ $bc/bc$-peak:\par
\centerline{
$\infer
    {\begin{aligned}
        \eq ~ \cup ~ \{ &V =_{}^? cons(v, Z), \; W =_{}^?
           cons(w, Z), \; U =_{}^? cons(u, U'), \; \\[-4pt]
        & U' =_{}^? bc(Z, u), \; u =_{}^? h(v, x), \; u =_{}^? h(w, y) \}
    \end{aligned}}
    {\eq ~ \uplus ~ \{ U =_{}^? bc(V, x), \; U =_{}^? bc(W, y) \}}$
}

\hsp ~~if $U \in \mathbf{nonnil}$

\vspace*{0.3em}
\lft (L5) {\em Splitting}, at a $cons/bc$-peak:\par
\centerline{
$\infer{\eq ~ \cup ~ \{ U =_{}^? cons(x, U_1^{}) , \;
                       V =_{}^? cons(y, V_1^{}), \;
             x =_{}^? h(y, z), \; U_1^{} =_{}^? bc(V_1^{}, x) \}}
        {\eq ~ \uplus ~ \{ U =_{}^? cons(x, U_1), \;
                        U =_{}^? bc(V, z) \}}$
}

\vspace*{0.3em}
\lft (L6) {\em Occur-Check Violation}:\\
\centerline{
$\infer[ \qquad \mathrm{if} ~ U ~ \mathrm{occurs ~ in} ~ \p, 
    \mathrm{, ~ and ~ }  {[}U{]} \succ_l {[}U{]}  ~
                   \mathrm{~ on ~ the ~ graph ~} G_l ]
    { FAIL } { \eq }$
}

\vspace*{0.3em}
\lft (L7) {\em Size Conflict}:\\
\centerline{
$\infer{ FAIL } { \eq ~ \uplus ~ \{ U =_{}^? cons(v, W),
        \; U =_{}^? nil \} }$
}

The symbol `$\uplus$' in the premises of the above inference rules 
stands for disjoint set union (and `$\cup$' for usual set union). 
The role of the Variable Elimination inference rule (L1) is to keep 
the propagation graph of $\p$ irredundant: each variable has a unique
representative node on $G_l(\p)$, up to variable equality. This rule 
is applied most eagerly. 
Rules~(L2), (L3.a)--(L3.c) and (L4.a) come next in priority, and then
(L4.b). The Splitting rule (L5) is applied in the ``laziest'' 
fashion, i.e., (L5) is applied only when no other rule is applicable.
The above inference rules  are all ``don't-care'' nondeterministic.  
(The priority notions just mentioned serve essentially for optimizing 
the inference procedure.) 

The validity of the rule (L4.b) (`Pushing $bc$ below $cons$') results 
from the cancellativity of $cons$ and the semi-cancellativity of $bc$ 
({\em Appendix-1}).  
Note that the variables $Z$, $U'$, and $u$ in the `inferred part' of this 
rule (L4.b) might need to be fresh; the same is true also for the variables 
$y$ and $V_2$ in the inferred part of the Splitting rule; but, in either case 
this is not obligatory, if the equations already present can be used for 
applying these rules. Type-inference failure is assumed to be 
checked implicitly; no explicit rule is given. 

The following point should be kept in mind: Any given problem $\p$ naturally 
`evolves' under the inference rules; and new variables might get added 
in the process, if rule (L5) or rule (L4.b) is applied; but none of the 
variables initially present in $\p$ can disappear in the process; not even 
under the Variable Elimination rule (L1). Thus, although the graph $G_l$ 
referred to in the Occur-Check Violation rule (L6) is the graph of the 
`current problem', the node it refers to might still be one corresponding 
to an initial variable.

\medskip
We show now that such an introduction of fresh variables cannot go for ever, 
and that the above ``don't-care'' nondeterministic rules suffice, 
essentially, for deciding {\em unifiability\/} modulo the axioms of $\bc$. 

\begin{prop}\label{list-unifiable} 
Let $\p$ be any $\bc$-unification problem, given in standard form.
The system $\Inf_l^{}$ of list-inference rules, given above, terminates 
on $\p$ in polynomially many steps.
\end{prop}

\noindent
\proof Assume given a problem $\p$ in standard form, for which the 
inference process does not lead to failure on Occur-Check (L6) or
Size-Conflict (L7).  
If  $\Inf_l^{}$ is  non-terminating on such a $\p$, at least one of the 
rules of $\Inf_l^{}$ must have been applied infinitely often along some 
inference chain; we show that this cannot be true for any of the 
rules in $\Inf_l^{}$.     

Note first that an equation of the form $U \Eq V$ in $\p$ 
is never handled in `both directions' by the variable elimination rule (L1); 
an application of this rule means: every occurrence of the variable $U$   
in the problem is replaced by the variable $V$.  
It is easy to check then, that for this reason, (L1) cannot give rise to 
non-termination. On the other hand, the list-inference rules (L2) through 
(L4.a) eliminate a (directed) outgoing arc from some node of $G_l$; so 
their termination is easy to check. It should be clear, that for these 
three rules, termination is polynomial (even linear). Thus, to show the 
termination of the entire inference process in polynomially many steps, 
we have to look at how the problem evolves under the rule~(L5) 
({\em Splitting}) and the rule~(L4.b) ({\em Pushing $bc$ below $cons$}).
We show that if occur-check violation (L6) does not occur, then the 
applications of the rule (L5) or of the rule (L4.b) cannot go on forever.

\medskip
For proving this, we shall be using an equivalence relation denoted 
as  ${\mathop{\sim}}_{\beta}^{}$, on the list-variables of the given problem.  
It is defined as the smallest equivalence relation\footnote{The relation
${\mathop{\sim}}_{\beta}^{}$ can be viewed as a combination of the {\em
unification closure\/}, a notion defined by Kanellakis and
Revesz~\cite{KanRev}, and the {\em congruence closure\/} of $\sim_{bc}^*$.  
The difference is that here we are working with a typed system.} satisfying 
the following conditions, on the list-variables of $\p$: 
\begin{itemize}
\item[-] If $U \sim_{bc}^* V$ then $U \mathop{\sim}_{\beta} V$.  
\item[-] Let $U\, >_{cons} \,U'$ and $V\, >_{cons} \,V'$; then 
$U  \mathop{\sim}_\beta  V$ implies $U'  \mathop{\sim}_\beta  V'$. 
\end{itemize}

\medskip
Observe now that the number of $bc$-equations, i.e.,
list-equations of the form $U =_{}^? bc(V, z)$, never increases. 
This number decreases in most cases, except for (L1), (L2)
and~(L5). The splitting rule~(L5) does not decrease
the number of $bc$-equations and may introduce new
variables, but the number of ${\mathop{\sim}}_{\beta}^{}$-equivalence 
classes of nodes (on the current graph) does not increase: 
Indeed, applying the splitting rule (L5) on a list-equation $U =^?  bc(V, z)$ 
removes that equation, and creates a list-equation of the form 
$U_1 =^? bc(V_1, x)$ for some list-variables $U_1$ and $V_1$, such 
that $V \sim_{bc}^{} U >_{cons} U_1 \sim_{bc}^{} V_1$; we have: 
$V_1^{} \sim_{\beta} U_1^{}$, since $V \sim_{\beta} U$.

Suppose now that applying the splitting rule does not terminate. Then, at 
some stage, the derived problem will have a sequence of variables of the 
form $U_0 >_{cons} U_1 >_{cons} \dotsb >_{cons} U_n$,  such that the length 
of the sequence $n$ strictly exceeds the initial number of 
$\sim_\beta$-equivalence classes --- which cannot increase under 
splitting, as we just observed above.  So there must exist indices 
$0 \le i < j \le n$ such that $U_i \sim_\beta U_j$.

Let $j \le n$ be the smallest integer for which there exists an
$i, \; 0 \le i < j$, such that  $U_i \sim_\beta U_j$. Then, by the definition 
of $\sim_\beta$, we must have $U_i \sim_{bc}^* U_j$. Consequently, we 
would then also have $[U_{i}] \succ_l [U_{i}]$; and that would have 
caused the inference process to terminate with FAIL, as soon as both the 
variables $U_i$ and $U_j$ appear in the problem derived under the inferences.  

Termination of (L4.b) can now be proved as follows: The number of 
$\sim^{*}_{bc}$-equivalence classes may increase by~1 with each application 
of~(L4.b), but the number of $\sim_{\beta}$-equivalence classes
remains the same, for the same reason as above. 
Let $m$ be the number of $bc$-equations in the input problem and let 
$n$ be the number of variables in the input problem. We then show 
that the total number of applications of (L4.b) and (L5) cannot
exceed~$mn$: Indeed, whenever one of (L4.b) or (L5) is applied, some number 
of $bc$-equations are removed and an equal or lesser number are added, whose 
variables belong to $\sim_\beta$-equivalence classes at a `lower  level' 
as explained above, i.e., below some $cons$ steps. There are at most $n$ 
such equivalence classes, since the number of $\sim_\beta$ equivalence 
classes does not increase (and there cannot be more than $n$ such 
equivalence classes, to start with).  
So a $bc$-equation cannot be ``pushed down'' more than $n$ times.
Since there are initially $m$ $bc$-equations, the total number of 
applications of (L4.b) and (L5) cannot exceed~$mn$. \qed

\medskip{}
A set of equations will be said to be {\em L-reduced\/} if none of the 
above inference rules (L1) through (L7) is applicable. (Note: such a problem 
may not be in $d$-solved form: an easy example is given a couple of paragraphs
below.) 

\medskip{}\noindent
{\bf Unification modulo $\bc$:} 
The rules (L1) through (L7) are not enough to show the existence of 
a unifier modulo $\bc$. The subset of element-equations, $\E(\p)$, may not 
be solvable; for example, the presence of an element-equation of the form 
$\{x =_{}^? h(x, z)\}$ should lead to failure. However, we have the following: 

\begin{prop}\label{list-result} 
If $\mathcal{L}(\p)$ is in L-reduced form, then $\p$ is unifiable modulo 
$\bc$ if and only if the set $\E(\p)$ of its element-equations 
is solvable.  
\end{prop}
\proof If $\mathcal{L}(\p)$ is $L$-reduced, then setting every list-variable
that is not in \textbf{nonnil} to $nil$ will lead to {\em a unifier\/}
for $\mathcal{L}(\p)$, modulo $\bc$, provided $\E(\p)$ is solvable. \qed

\medskip{}
Recall that $\bc_0^{}$ is the theory defined by $\bc$ when $h$ is uninterpreted. 
\begin{prop}\label{list-poly} 
Let $\p$ be any $\bc_0^{}$-unification problem, given in standard form.
Unifiability of $\p$ modulo $\bc_0^{}$ is decidable in polynomial time (wrt 
the size of  $\p$).
\end{prop}

\proof If the inferences of $\Inf_l^{}$ applied to $\p$ lead to failure, 
then $\p$ is not unifiable modulo  $\bc$; so assume that this is not the case, 
and replace $\p$ by an equivalent problem which is $L$-reduced, deduced 
in polynomially many steps by Proposition~\ref{list-unifiable}. 
By Proposition~\ref{list-result}, the unifiability modulo $\bc$ of such 
a $\p$ amounts to checking if the set $\E(\p)$ of its element-equations is 
solvable. We are in the case where $h$ is uninterpreted, so to solve $\E(\p)$ 
we apply the rules for standard unification, and check for their termination  
without failure; this can be done in polynomial time~\cite{BaaderSnyd-01}.  
(In this case, $h$ is fully cancellative.) \qed 

It can be seen that while termination of the above inference rules
guarantees the {\em existence\/} of a unifier (provided the element 
equations are syntactically solvable), the resulting $L$-reduced
system may not lead directly to a unifier. For instance, the $L$-reduced 
system of list-equations $\{ U =_{}^? bc(V, x), \; U =_{}^? bc(V, y) \}$ 
is unifiable, with the following two incomparable unifiers:  
\disp{$\{ x := y, \, U := bc(V, y) \} \; \mathrm{~ and ~}
                \{ U := nil, \, V := nil \}$} 

To get a complete set of unifiers we need three more inference rules,  
which are ``don't-know'' nondeterministic, to be applied only to 
$L$-reduced systems: 

\medskip{}
\lft (L8) {\em Nil-solution-Branch for $bc$\/}, at a $bc/bc$-peak:\par
\centerline{
$\infer
{\eq ~ \cup ~ \{ U =_{}^? nil, \; V =_{}^? nil, \; W =_{}^? nil \}}
   {\eq ~ \uplus ~ \{ U =_{}^? bc(V, x), \; U =_{}^? bc(W, y) \}}$
}

\medskip{}
\lft (L9) {\em Guess a non-Nil branch for $bc$\/}, at a $bc/bc$-peak:\par

\centerline{
$\infer
    {\begin{aligned}
        \eq ~ \cup ~ \{ &V =_{}^? cons(v, Z), \; W =_{}^?
           cons(w, Z), \; U =_{}^? cons(u, U'), \; \\[-4pt]
        & U' =_{}^? bc(Z, u), \; u =_{}^? h(v, x), \; u =_{}^? h(w, y) \}
    \end{aligned}}
    {\eq ~ \uplus ~ \{ U =_{}^? bc(V, x), \; U =_{}^? bc(W, y) \}}$
}

\medskip{}
\lft (L10) {\em Standard Unification on $bc$}:\par

\centerline{
$\infer
{\eq ~ \cup ~ \{ U =_{}^? bc(W, y), \; V =_{}^? W, \; x =_{}^? y \}}
{\eq ~ \uplus ~ \{ U =_{}^? bc(V, x), \; U =_{}^? bc(W, y) \}}$
}

\medskip{}
Rule (L9) nondeterministically `guesses' $U$ to be in {\bf nonnil};  
in other words, it applies rule (L4.b) `unconditionally'. 
The inference system thus extended will be referred to as $\Inf_l'$. 
By the same reasonings as developed above, $\Inf_l'$ also terminates, in
polynomially many steps, on any problem given in standard form. 
We establish now a technical result, valid whether or not $h$ 
is interpreted: 

\medskip{}
\begin{prop}~\label{d-solved}
Let $\p$ be any $\bc$-unification problem in standard form, to which 
none of the inferences of $\Inf_l'$ is applicable. 
Then its set of list-equations is in $d$-solved form. 
\end{prop}
\proof
If none of the equations in $\p$ involve $bc$ or $cons$ (i.e., all equations
are equalities between list-variables), then the proposition is proved 
by rule (L1) ({\em Variable Elimination}).

Observe first that if $\Inf_l$ is inapplicable to $\p$, then, on the 
propagation graph $G_l$ for $\p$, there is {\em at most one outgoing directed 
arc} of $G_l$ at any node $U$: Otherwise, suppose there are two distinct 
outgoing arcs at some node $U$ on $G_l$; if both directed arcs bear the 
label $>_{cons}$, then rule (L2) of $\Inf_l$  would apply; if both bear the 
label $>_{bc}$, then one of (L4.a), (L4.b), (L9), (L10)  would apply; the 
only remaining case is where one of the outgoing arcs is labeled with  
$>_{cons}$ and the other has label~$>_{bc}$, but then the splitting  
rule~(L5) would apply. 

Consider now any given connected component $\Gamma$ of $G_l$. There can be 
no directed cycle from any node $U$ on $\Gamma$ to itself: otherwise the 
Occur-Check-Violation rule (L6) would have applied. It follows, from this 
observation and the preceding one, that there is a 
unique {\em end-node\/} $U_0$ on $\Gamma$, i.e., a node from which there 
is {\em no directed outgoing arc\/}; and also that for any given node 
$U$ on $\Gamma$, there is a unique well-defined directed path leading 
from $U$ to that end-node $U_0$.   

It follows easily from these, that the list-variables on the left hand sides 
of the equations in~$\p$ (on the different connected components of $G_l$) can 
be ordered suitably, so as to satisfy the condition for $\p$ to be in  a 
$d$-solved form. \qed 

\begin{exa}  
The following $\bc_0^{}$-unification problem is in standard form: 
\disp{$U \Eq cons(x, W), \; ~U  \Eq bc(V, y), \; ~W \Eq bc(V_2, y), \;
            x \Eq h(z, y), \;  ~ y \Eq a$} 
We apply (L5) ({\em Splitting}) and write $V \Eq cons(v_1, V_1)$,
with $v_1, V_1$ fresh; this, followed by an application of rule (L2) 
({\em Cancellation on cons}) leads  to: 
\disp{$U \Eq cons(x, W), \; V \Eq cons(v_1, V_1), \; 
           W \Eq bc(V_1, x), \; W \Eq bc(V_2, y), \;$ \\
         $x \Eq h(v_1, y), \; x \Eq h(z, y), \; y \Eq a$}

\lft
We apply cancel\-lativity of $h$ (valid for $\bc_0$), and an element-variable 
elimination; the problem thus derived is the following:
\disp{$U \Eq cons(x, W), \; V \Eq cons(z, V_1), \; 
          W \Eq bc(V_1, x), \; W \Eq bc(V_2, y), \;$ \\
            $x \Eq h(v_1, y), \; z \Eq v_1, \; y \Eq a$}

\lft
(i) No rule of $\Inf_l$ is applicable: in particular, (L4.b) doesn't apply 
since $W$ is not in {\bf nonnil}; but the rule (L8) ({\em Nil-solution 
Branch for $bc$}) can be nondeterministically applied:

\disp{$U \Eq cons(x, W), \; W \Eq nil, \; V_1 \Eq nil, \; V_2 \Eq nil, \;
            V \Eq cons(z, V_1),$ \\
            $x \Eq h(v_1, y), \; z \Eq v_1, \; y \Eq a$}

These equations, in $d$-solved form, give a solution to the original 
problem.  

\medskip{}\lft
(ii) For the sake of completeness, we could also try the rule (L9) 
({\em Guess a non-Nil branch}) nondeterministically, 
successively on the two equations for $W$ in the problem derived above; 
so we write $V_1 =^? cons(v_2, V'_2)$ and  $V_2 =^? cons(v_3, V'_3)$. 
These applications of (L9), followed by applications of {\em Variable 
elimination, Cancellation on cons}, and the  cancellativity of $h$ 
(valid for the theory $\bc_0$), will lead us to:   

\disp{$U \Eq cons(y, W), \; V \Eq cons(v_1, V_1), \;  
                  V_1 =^? cons(v_3, V'_3),  \;$ 
         $V_2 =^? V_1,  V'_2 =^? V'_3, \;  W \Eq bc(V_1, x), \;$  \\
     $x \Eq y, \; y \Eq h(v_1, y), \; v_2 \Eq v_3, \; z \Eq v_1, \; y \Eq a$}

\lft
The list-equations are in $d$-solved form, but the element-equations 
being unsatisfiable we are led to failure.
\enlargethispage{\baselineskip} 

\medskip{}\lft
(iii) For the following problem (almost same as (i) above, but 
for an element-equation): 
\disp{$U \Eq cons(x, W), \; ~U  \Eq bc(V, y), \; ~W \Eq bc(V_2, y), \;
            ~ y \Eq a$} 
the reasonings as developed in (ii) above would have led us to a 
non-nil solution for $W$: 
\disp{$U \Eq cons(y, W), \; V \Eq cons(v_1, V_1), \;  
                  V_1 =^? cons(v_2, V'_3),  \;$  
         $V_2 =^? V_1, \;  W \Eq bc(V_1, x), \;$  
             $x \Eq y, \;  \; y \Eq a$}
where $V_3'$ is any arbitrary list, and $v_1, v_2$ are any arbitrary 
elements.                  \qed 

\end{exa}

We turn our attention in the following section to the unification problem 
modulo $\bc$. When $h$ is uninterpreted, we saw that this unification is 
decidable in polynomial time. But when $h$ is interpreted so that $\bc$ 
models CBC, we shall see that unification modulo $\bc_1^{}$ is NP-complete.

\section{Solving a $\bc$-Unification problem}\label{method}
 
Let $\p$ be a $\bc$-Unification problem, given in standard form. We assume 
that $\Inf'_l$ has terminated without failure on $\p$; we saw, in the
preceding section (Proposition~\ref{d-solved}), that $\p$ is then in 
$d$-solved form. We also assume that we have a sound and complete procedure 
for solving the element-equations of $\p$, that we shall denote as 
$\Inf_e$. For the theory $\bc_0^{}$ where $h$ is uninterpreted, 
we know (Proposition~\ref{list-poly}) that $\Inf_e$ is standard 
unification, with cancellation rules for $h$, and failure in case of 
`symbol clash'. For the theory $\bc_1^{}$, where $h(x, y)$ is interpreted 
as $e_k(x\oplus y)$ for some fixed key $k$, $\Inf_e$ will have rules for 
semi-cancellation on $h$, besides the rules for unification modulo XOR in 
some fixed  procedure; such a procedure is assumed given once and 
for all. 

In all cases, we shall consider $\Inf_e$ as a black-box that either 
returns most general unifiers ({\em mgu\/}'s) for the element-equations of 
$\p$, or a failure message when these are not satisfiable. Note that  
$\Inf_e$ is unitary for $\bc_0$ and finitary for $\bc_1$. 
For any  problem  $\p$ in $d$-solved form,  satisfiable under the theory 
$\bc_0$, there is a unique mgu, as expressed by the equations of $\p$ 
themselves (cf. also \cite{JoKi}), that we shall denote by $\theta_{\p}$. 
Under $\bc_1^{}$ there could be more than one (but finitely many) mgu's; 
we shall agree to denote by $\theta_{\p}$ any one among them.   
The entire procedure for solving any $\bc$-unification problem $\p$, 
given in standard form, can now be synthesized as a nondeterministic 
algorithm:  

\medskip{}
\noindent
{\bf The Algorithm $\A$:} Given a $\bc$-unification problem $\p$, in  
standard form. \\  
 \hsp $G_l$ = Propagation graph for $\p$. \\
 \hsp $\Inf_l'$ =  Inference procedure given above for $\mathcal{L}(\p)$. \\
 \hsp $\Inf_e^{}$ = Any given (complete) procedure for solving 
   the equations of $\E(\p)$. 

\begin{enumerate}
\item Compute a standard form for $\p$, to which the ``don't-care''  
inferences of $\Inf_l^{}$ are no longer applicable. If this leads to 
failure, exit with FAIL. Otherwise, replace $\p$ by this standard form. 

\item Apply the ``don't-know'' nondeterministic rules (L8)--(L10), 
followed by the rules of $\Inf_l^{}$ as needed, until the equations 
no longer get modified by the inference rules (L1)--(L10).  If this leads to 
failure, exit with FAIL. 

\item Apply the procedure $\Inf_e$ for solving the residual set $\E(\p)$ 
of element-equations; if this leads to failure, exit with FAIL. 

\item Otherwise let $\sigma$ be the substitution on the variables 
of $\p$ as expressed by the resulting equations. Return $\sigma$ as 
a solution to $\p$.
\end{enumerate}

\medskip{}
\begin{prop}~\label{complete}
The algorithm $\A$ is sound and complete. 
\end{prop}
\proof The soundness of $\A$ follows from the soundness (assumed) of 
$\Inf_e$ and that of $\Inf_l'$, which is easy to check: obviously,  
if $\p'$ is any problem derived from $\p$ by applying any of these 
inference rules, then any solution for $\p'$ corresponds to a solution 
for $\p$. The completeness of $\A$ follows from the completeness (assumed) 
of $\Inf_e$, and the completeness of $\Inf_l'$ that we prove below. \qed

\medskip{}
\begin{lem}~\label{soln-preserve}
If $\sigma$ is a solution for a given $\bc$-unification problem $\p$ 
in standard form, then there is a sequence of $\Inf_l'$-inference 
steps that transforms $\p$ into a problem $\p'$ in $d$-solved form 
such that $\sigma$ is an instance of $\theta_{\p'}^{}$ (modulo $\bc$).
\end{lem}

\noindent
\proof We know that the inference rules of $\Inf_l'$ terminate on $\p$;   
let $N$ be the maximum number of steps needed for this termination, 
including along all possible ``don't-know'' branches of the process. 
We prove the lemma  by induction on $N$, and case analysis for the 
possible branches. 

Observe first that if $\p'$ is a problem derived from $\p$ under 
any inference rule of $\Inf_l'$, then the given substitution $\sigma$, on  
on the variables of $\p$, extends naturally as a substitution on the 
variables of $\p'$, satisfying the equations of $\p'$.  
(This needs to be checked only if $\p'$ might involve new variables, such 
as when $\p'$ is derived from $\p$ under rule (L5) or rule (L4.b); 
 the reasoning is straightforward for either of these cases.)  

If $\p'$ is derived from $\p$ by applying one of the ``don't-care'' rules 
of $\Inf_l$, then  the assertion of the lemma follows from the above
observation and the induction hypothesis. So we may assume wlog that the 
given problem $\p$ is already $L$-reduced (i.e., none of the inferences of 
$\Inf_l$ is applicable).  If such a $\p$ is already in $d$-solved form, then 
we are done, since $\sigma \, \preceq_{\bc} \, \theta_{\p}$, for some mgu
$\theta_{\p}$. (If the theory is $\bc_1$, this means:  there exists one 
among the finitely many $mgu$s, for which this holds.) 

If $\p$ is not in $d$-solved form, then several cases are possible, 
depending on the possible inference branches. 
It suffices to consider one such case -- the reasoning being quite similar 
for all the others. Suppose there are two equations 
$U =_{}^? bc(Z, v)$ and $U =_{}^? bc(Y, w)$ in $\p$. 
If $\sigma (v) =_{\bc}^{} \sigma (w)$, then we must have 
$\sigma (Z) =_{\bc}^{} \sigma (Y)$, and $\sigma$ is extendable as a 
solution for the problem obtained by applying the rule~(L10). If $\sigma (v) 
\neq_{\bc}^{} \sigma (w)$, then $\sigma$ must be extendable as a solution to 
the problem derived under rule (L8) or rule (L9). The induction hypothesis 
(on the maximum number of inference steps needed for termination) 
completes then the argument to prove the lemma, in all cases.  \qed

\medskip{}
\begin{prop}~\label{finitariness}
Unification modulo $\bc$ is finitary.
\end{prop}
\noindent
\proof  Let $\p$ be a satisfiable $\bc$-unification problem. We can assume 
without loss of generality that $\p$ is in standard form, because any 
unification problem can be converted to a finite problem in standard form. 
Let $S$ be the set of mgus for $\p$.  By lemma~\ref{soln-preserve}, for 
each $\sigma \in S$, there is a sequence of $\Inf_l'$-inference steps that 
leads to a problem $\p'$ in $d$-solved form, and an $mgu$ $\theta_{\p'}$  
such that $\sigma$ is an instance of $\theta_{\p'}$. 
Let $D$ be the set of all such derived problems. Because all the inference 
rules in $\Inf_l'$ terminate, and because there are finitely many 
inference rules, $D$ contains finitely many problems.

In the uninterpreted case $\bc_0$, $\sigma$ is $\theta_{\p'}$ for some $\p' 
 \in D$, so there are finitely many unifiers in $S$. For $\bc_1$, note that
unification modulo XOR is finitary~\cite{clynch-liu}. Therefore, there are
finitely many XOR-mgus for the element problem derived from $\p'$, so there 
are finitely many unifiers in $S$ that are instances of $\theta_{\p'}$. 
Since there are finitely many problems in $D$, there are finitely many 
unifiers in $S$. \qed

\subsection{$\bc_1$-Unification is NP-Complete}\label{complex}

Recall that $\bc_0$ is the theory defined by $\bc$ when $h$ is uninterpreted,
and $\bc_1$ is the theory when $h$ is interpreted so that $\bc$ models the
(XOR-based) cipher-block-chaining mode CBC. 

\begin{prop}
Unifiability modulo the theory $\bc_1^{}$ is NP-complete. 
\end{prop}
\proof NP-hardness follows from the fact that general unification modulo 
XOR is NP-complete~\cite{GuoDranWolf-Cade96}. 
We deduce the NP-upper bound from the following facts: 

\begin{enumerate}[label=\alph*)]
\item  For any given  $\bc$-unification problem, computing a standard
   form is in polynomial time, wrt the size of the problem.
 
\item Given a standard form, the propagation graph can
   be constructed in polynomial time (wrt its number of
   variables). 

\item Applying (L1)-(L10) till termination takes only polynomially
   many steps.
 
\item Extracting the set of element-equations from the resulting 
set of equations is in P. 

\item Solving the element-equations, with the procedure $\Inf_e$, using
   unification  modulo XOR, is in NP. \qed
\end{enumerate}

\subsection{An Illustrative Example}~\label{ill-examp}

\medskip{}
The following public key protocol is a slight variant of one that was 
studied in~\cite{DolevEvenKarp} --  the modification is that the 
namestamp of the sender of a message appears as the {\em first} block 
of the encrypted message body, and not the second as was specified 
in~\cite{DolevEvenKarp}:  

\medskip\lft
\hspace*{3cm} $A \rightarrow B: A, \, \{A, m \}_{kb}^{}$ \\
\hspace*{3cm} $B \rightarrow A: B, \, \{B, m \}_{ka}^{}$ \\
where $A, B$ are the participants of the protocol session, $m$ is a  
message that they intend secret for others, and $kb$ (resp. $ka$) 
is the public key of $B$ (resp. $A$). 

If the CBC encryption mode is assumed and the message blocks are 
all of the same size, then this protocol becomes insecure; here is why.  
Let $e_Z(x)$ stand for the encryption $e(x, kz)$ with the public key $kz$ of 
any principal $Z$. Under the CBC encryption mode, what $A$ sends to $B$ 
is the following list, in the ML-notation: \\
\hspace*{2cm} $A \rightarrow B: 
    [\, A, \,[\, e_B(A\oplus v), \; e_B(m \oplus e_B(A \oplus v))\, ]\,]$. \\
Here $\oplus$ stands for XOR and $v$ is the initialization vector ($IV$)
agreed upon between $A$ and $B$.  But then, some other agent $I$, entitled 
to open a session with $B$ with initialization vector $w$, can get hold of 
the first encrypted block (namely: $e_B(A\oplus v)$) as well as the second 
encrypted block of what $A$ sent to $B$, namely $e_B(m, e_B(A\oplus v))$; 
(s)he can then send the following as a `bona fide' message to $B$: 

\hspace*{7mm} 
 $I \rightarrow B: 
  [\, I, \,[\, e_B(I \oplus w), \; e_B(m \oplus e_B(A\oplus v))\,] \, ]$; \\
upon which $B$ will send back to $I$ the following: \\ 
 \hspace*{12mm} 
$B \rightarrow I: [\,B, \, [ \, e_I(B\oplus w), \;
 e_I(\, m \oplus e_B(A\oplus v)  \oplus e_B(I\oplus w) \oplus e_I(B\oplus w) 
             \, ) \,] \, ]$. 

\medskip
It is clear now, that the intruder $I$ can get hold of the message 
$m$ intended to remain  secret for him/her: 
By decrypting the second block of the (encrypted part of the) message  
received from $B$, (s)he first deduces: 
 $ m \oplus e_B(A\oplus v)  \oplus e_B(I\oplus w) \oplus e_I(B\oplus w) $; 
by XOR-ing this with the first block of the message, (s)he obtains:  
 $ m \oplus e_B(A\oplus v)  \oplus e_B(I\oplus w) $; from which (s)he can 
deduce $m$ by XOR-ing with $e_B(I\oplus w)$ and  $e_B(A\oplus v)$,  
both of which are known to him/her (the latter of these two terms is the 
first block of the message from $A$ to $B$, that (s)he has intercepted).   

\medskip{}
\begin{exa}~\label{ping} 
The above attack (which exploits  the properties of XOR: 
$x \oplus x = 0, ~x \oplus 0 = x$) can be modeled as solving 
a certain $\bc_1^{}$-unification problem. We assume that  the names 
$A, B, I$,  as well as the initialization vector $w$, are  constants 
accessible to $I$. The message $m$  and the initialization vector $v$, 
that $A$ and $B$ have agreed upon, are constants intended to be secret for 
$I$. We shall interpret the function symbol $h$ of $\bc$ in terms of encryption 
with the public key of $B$: i.e., $h(x, y)$ is $e_B(x\oplus y)$. 

The protocol above  can then be modeled as follows: We assume that the list 
of terms $A$ sends to $B$, namely $[\, A, [h(A, v), h(m, h(A, v))] \,]$, is
seen by the latter as the list of terms  $[\, A, bc([A, m], v)\, ]$; (s)he 
first recovers the namestamp $A$ of the sender, then checks that the second 
argument under $bc$ in what (s)he received is the $IV$ agreed upon with $A$; 
subsequently (s)he  sends back the appropriate list of terms to 
$A$, acknowledging receipt of the message.  

Now, due to our CBC-assumption, the ground terms $h(A, v), \; 
h(m, h(A, v))$ are both accessible to the intruder $I$. So the attack by $I$,  
mentioned above, corresponds to the fact that $I$ {\em can\/} send to $B$ 
the following list of terms: $[\, I, [\, h(I, w), h(m, h(A, v))\, ]\,]$. 
That the attack materializes follows from the fact that $B$ can solve the 
$\bc_1$-unification problem:
\disp{$bc([I, z], w) \Eq cons(h(I, w), [h(m, h(A, v))])$, }
for the element-variable $z$, i.e., $B$ needs to solve the 
element-equation: $h(z,  h(I, w)) \Eq  h (m, h(A, v))$; since $h$ is 
interpreted here so that $\bc$ models $CBC$, (s)he can do so by setting: 
$z := m \oplus h(A, v) \oplus h(I, w)$; and that precisely leads to 
 the attack.     \qed 
 
\end{exa} 

\begin{rem}\label{R:1}
(i) The above analysis does {\em not\/} go through if the namestamp forms 
the {\em second block\/} of the encrypted part of the messages sent. In 
such a case, the protocol is `leak-proof' even under CBC, provided we assume 
that an IV for a message is a secret to be shared only by the sender and 
the intended recipient of the message, and that it is {\em not} 
transmitted -- as clear text or encrypted -- as an initial `block number 
 zero' of the message body. 
Actually, by reasoning as above, one checks that the intruder $I$ 
in such a case can only get hold of $m\oplus v$, where $v$ is the (secret) 
IV that only $A$ and $B$ share. This in a sense is in accordance 
with~\cite{DolevEvenKarp}, where the protocol was `proved secure'  
under such a specification. 

\medskip{}
(ii) The considerations above lead us to conclude, implicitly, that 
in cryptographic protocols employing the CBC encryption mode, it is 
necessary to forbid free access to the IVs of the `records' of the 
`messages' sent, if information leak is to be avoided. This fact has 
been pointed out in the 90's, by Bellare et al (\cite{Rogaway1995}), and 
again, in some detail, by K.~G.~Paterson et al in \cite{paterson2011}; 
both point out that TLS~1.0 -- with its predictable IVs -- is inherently 
insecure. 
For more on this point, and on the relative advantages of TLS~1.1, 
TLS~1.2 over TLS~1.0, the reader can also consult, e.g.,  
 {\tt http://www.educatedguesswork.org/2011/09/}

(Note: keeping IVs as shared secrets alone may not always be sufficient  
in general, as is shown by Example 2 above.) 
\end{rem}

\section{A generic Block Chained Cipher-Decipher Scheme}~\label{DBC}

In this section we extend the 2-sorted equational theory $\bc_0$ studied 
above, into one that fully models, in a simple manner and without using 
any AC-symbols, a `generic' block chaining encryption-decryption scheme. 
This theory, that we shall refer to as $\dbc$, is defined by the following 
set of (2-sorted) equations: 
\begin{eqnarray*}
bc(nil, \, z) & = & nil \\
bc(cons(x, Y), \, z) & = & cons(h(x,z), \; bc(Y, \, h(x,z)))\\
g(h(x,y), \, y) & = & x \\
db(nil, \, z) & = & nil \\ 
db(cons(x, Y), \, z) & = & cons(g(x, z), \; db(Y, \, x)) \\
db(bc(X, y), \, y) & = & X 
\end{eqnarray*}
where $g$ is typed as $g: \; \tau_e  \times \tau_e \rightarrow \tau_e$ 
and $db$ is typed as $db: \; \tau_l \times \tau_e \rightarrow \tau_l$. 

\medskip{}
All these equations can be oriented from left to right under a suitable 
reduction ordering, to form a convergent (2-sorted) rewrite system. 
The $6$th equation says that $db$ is a left-inverse for 
$bc$; it is actually an inductive consequence of the first five: i.e., for 
any list-term $X$ and element-term $y$ both in ground normal form, 
$db(bc(X, y), y)$ reduces to $X$ under the first five, a fact that can 
be easily checked by structural induction, cf. {\em Appendix-2}.   
(Its insertion as an equational axiom is for technical reasons, as will be  
explained in {\em Remark \ref{R:4}}(ii) below.) 

A few words, by way of intended semantics in the context of cryptographic 
protocols, seem appropriate: $h(x, y)$ would in such a context 
stand for the encryption with the public key of an intended recipient $B$, 
of message $x$, {\em `coupled' in a sense to be defined,} with $y$ as 
initialization vector (IV); and $g(h(x,y), \, y)$ would be the 
decryption of $h(x, y)$ with the private key of $B$, to be then 
 {\em `decoupled', again in a sense to be defined}, with $y$.  
If an agent $A$ wants to send a list of terms $cons(x, Y)$ to recipient 
$B$, (s)he would send out $bc(cons(x, Y), \, z)$ where $z$ is the IV 
they have mutually agreed upon; and $B$ would see it as the list of 
terms $cons(h(x,z), \; bc(Y, \, h(x,z)))$, from which (s)he can retrieve 
the individual message terms by applying the last equation for $db$ in 
the system $\dbc$.  

This generic block chained encryption-decryption scheme is a natural 
abstraction of the usual (XOR-based) CBC: it suffices to interpret the 
roles of $h$ and $g$ suitably, and define properly the meanings of `coupling' 
and `decoupling', to get the usual CBC mode; for that, one would {\em define} 
the `coupling' as well as `decoupling' of $x$ with $y$ as $x \oplus y$;  
$h(x, y)$ would then stand for $e_B(x \oplus y)$, and $g(z, y)$ would 
stand for  $d_B(z) \oplus y$, where $d_B$ is  decryption with the private 
key of $B$. If we go back to Example~\ref{ping} based on  the usual CBC, 
the encrypted part of what $A$ sends out to $B$ (with the notation employed 
there) is the list of terms: $[ \, h(A, v), \; h(m, h(A, v)) \, ]$, 
that corresponds to the term $bc([A, m], v)$. 
By applying the fifth equation in $\dbc$ to this list of terms, 
under the assignments: $z:= v, \, x := h(A, v), \, Y := [h(m, h(A, v)]$,  
$B$ would then derive the following list:  
\disp{$[ \, g(h(A, v), v), \; db( [h(m, h(A, v))], h(A, v) )\,]$;} 
i.e., the list $[A, m]$.  In other words, the usual XOR-based 
CBC is indeed an `instance' of the theory $\dbc$.  

\medskip{}
\begin{rem}\label{R:2} Other `concrete' cipher-decipher block chaining modes 
can also be seen  as instances of $\dbc$; one among them is the 
{\em Cipher FeedBack encryption mode\/} (CFB), which is defined as follows: 

Let $M = p_1  \dots p_n$ be a message given as a list of $n$ `plaintext' 
message subblocks. Then the encryption of $M$ with any given key $k$ and 
initialization vector $v$ is defined as the list $c_1 \dots c_n$, of 
ciphertext message subblocks, where: 
\disp{$c_1 = p_1 \oplus e_k(v)$, \, and ~~$c_i = p_i \oplus e_k(c_{i-1})$, 
     for any  $1 < i \le n$}

This encryption mode (also using XOR) is very similar to CBC, 
but  works in the reverse direction (cf. e.g., 
{\tt http://en.wikipedia.org/wiki/Block\_cipher\_modes\_of\_operation}). 
It is an instance of $\dbc$, if the `coupling' and the `decoupling' 
operations of $\dbc$, namely $h(x, y)$ and $g(x, y)$, are both defined as 
$x \oplus e_k(y)$.
\end{rem}

\medskip{}
The theory $\dbc$ thus appears, indeed, as a high level equational 
abstraction of the block chained encryption-decryption mode; it employs  
no AC-symbols for this abstraction.  It is easy to see, on the other hand, 
that the equations of $\dbc$ can all be oriented left-to-right under a 
suitable reduction ordering, to give a convergent rewrite system. 
We shall be showing below that unification modulo  $\dbc$ is NP-decidable; 
it turns out to be actually NP-complete, due to the presence of a 
left-inverse for $h$ (namely $g$).  

\medskip{}
\begin{rem}\label{R:3}:  It is important to note that the function~$g$ is not
semi-cancellative: $g(h(g(t,u), u), u) =_{\dbc}^{} g(t,u)$, 
but $h(g(t,u), u)$ and $t$ need not be equivalent modulo~$\dbc$.
However, it is easy to show that $g$ is left-cancellative; see 
{\em Appendix-1} for the details.   
\end{rem}

\subsection{Unification modulo $\dbc$}~\label{unifmod-dbc}

\medskip
We assume without loss of generality that any $\dbc$-unification
problem $\p$ is given in a standard form, i.e., as a set of equations
$\eq$, each having one of the following forms:
\disp{$U =_{}^? V, \; U =_{}^? bc(V, y),\; U =_{}^? db(V, y),\; 
       U =_{}^? cons(v, W),  \; U =_{}^? nil, \;$ \\
   $u =_{}^? v, \; u=_{}^? g(w,y), \; v=_{}^? h(w,x), \; u \Eq const$}

We have to extend some of the notions and notation of Section~\ref{Inf-l}, 
in order to take~$db$ into account.  These extensions concern the propagation 
graph $G_l$ of the problem and {\bf nonnil}, the set of variables which 
cannot be {\em nil.\/}

\begin{enumerate}[label=(\roman*)]
\item If $U =^? db(V, y)$ is in $\p$, then write $U  >_{db}  V$; in which
  case, insert a directed arc on $G_l$ from  $[U]$ to $[V]$ and label it 
  with  $>_{db}$.
  The graph $G_l$ will also have then a {\em two-sided  (undirected)\/} edge
  between $[U]$ and $[V]$, labeled with $\sim_{db}$. 

\item The set of variables  {\bf nonnil}, defined earlier, is extended 
as follows:  \\ If $U =^? db(V, y)$ is in $\p$, then $U$ is in {\bf nonnil} if
and only if $V$ is in {\bf nonnil}.
\end{enumerate}

\medskip
\noindent We define a new relation $>_c \, = \, >_{bc} \cup >_{db}$. Its symmetric
closure is $\sim_c$ and its transitive, reflexive, and symmetric closure 
is $\sim_c^*$. The relations $>_c^{+}, \, >_{db}^{+}, \, >_{db}^{*}$ are then 
defined in the usual manner. If $U \sim_c V$, then $U$ and $V$ are 
related by `chaining', i.e. by some number of $bc$ and $db$ operations.
We refine then the partial relation $\succ_l$ on the nodes of $G_l$ 
as follows:  
\disp{$\succ_l ~ = ~ \sim_c^* \; \circ \; >_{cons} \, \circ \;
           (\sim_c \cup >_{cons})^* $}
This relation can still continue to be read as: $[U] \succ_l [V]$ iff there 
is a directed path on $G_l$ from $[U]$ to $[V]$, at least one arc of which 
has label $>_{cons}$. 

\medskip
We extend now the inference system $\Inf'_l$ of Section~\ref{Inf-l} by 
adding the following list-inferences; these additional rules are essentially 
the $db$-counterparts of the list-inferences  of $\Inf'_l$ which only needed
to consider $bc$. (There are several reasons why we have not worked 
with $\dbc$ right from the start -- maybe the inference system would 
possibly have been more concise, if we had done so.  
A first reason is, that would have been at the expense of readability; 
a second reason is that $\bc$-unification is of interest on its own, 
especially for $\bc_1$, as is shown by Example~\ref{ping} above; 
a third and conclusive reason is that the inference system we present 
below for $\dbc$-unification, actually reduces the problem to a 
problem of $\bc$-unification.) 
We first formulate the  ``don't-care'' nondeterministic inference rules. 

\medskip
\begin{description}[font=\normalfont]
    \item[(DB1.a) {\em Nil solution-1 for $db$}:]
        \[ \infer
        {\eq ~ \cup ~ \{ \;  U =_{}^? nil, \; V =_{}^? nil  \; \}}
        {\eq ~ \uplus ~ \{ \;  U =_{}^? db(V, x), \; U =_{}^? nil  \; \}} \]

    \item[(DB1.b) {\em Nil solution-2 for $db$}:]
        \[ \infer
        {\eq ~ \cup ~ \{ \;  U =_{}^? nil, \; V =_{}^? nil  \; \}}
        {\eq ~ \uplus ~ \{ \;  U =_{}^? db(V, x), \; V =_{}^? nil  \; \}} \]

    \item[(DB1.c) {\em Nil solution-3 for $db$}:]
        \[ \infer[\qquad \mathrm{if} ~ V >_{db}^{*} U ]
        {\eq ~ \cup ~ \{ \;  U =_{}^? nil, \; V =_{}^? nil  \; \}}
        {\eq ~ \uplus ~ \{ \;  U =_{}^? db(V, x)  \; \}} \]

    \item[(DB2) {\em Left-Cancellation on $db$}:]
        \[ \infer[\quad \mathrm{if} ~ U \in \mathbf{nonnil}]
        {\eq ~ \cup ~ \{ \; U =_{}^? db(V, y), \; x =_{}^? y  \; \}}
        {\eq ~ \uplus ~ \{ \;  U =_{}^? db(V, x), \; U =_{}^? db(V, y)  \; \}} \]

    \item[(DB3.a) {\em Push $db$ below $cons$, at a 
                           $\mathbf{nonnil}$ $db/db$-peak }:]
        \[ \infer
        { \begin{aligned}\eq ~ \cup ~ \{ \; 
            & V =_{}^? cons(v, V'), \; W =_{}^? cons(w, W'), \;  
                       U =_{}^? cons(u, U'), \\[-4pt]
            & U' =_{}^? db(V', v), \; U' =_{}^? db(W', w), \; 
                       u =_{}^? g(v, x), \; u =_{}^? g(w, y)  \; \}
        \end{aligned} }
        {\eq ~ \uplus ~ \{ \;  U =_{}^? db(V, x), \; U =_{}^? db(W, y) \;
          \}} \]

       \hspa ~~${if} ~ U \in \mathbf{nonnil}$

    \item[(DB3.b) {\em Push $bc$ and $db$ below $cons$ at a 
                                $\mathbf{nonnil}$ $bc/db$-peak }:]
        \[ \infer
        { \begin{aligned}\eq ~ \cup ~ \{ \; 
            & V =_{}^? cons(v, V'), \; W =_{}^? cons(w, W'), 
                           \;  U =_{}^? cons(u, U'), \\[-4pt]
            & U' =_{}^? bc(V', u), \; U' =_{}^? db(W', w), 
                          \; u =_{}^? h(v, x), \; w =_{}^? h(u, y)  \; \}
        \end{aligned} }
        {\eq ~ \uplus ~ \{\; U =_{}^? bc(V, x), \; U =_{}^? db(W, y) \;\}} \]

     \hspa ~${if} ~ U \in \mathbf{nonnil}~$

    \item[(DB4) {\em Splitting for $db$ at a $cons/db$-peak}:]
        \[ \infer
        {\eq ~ \cup ~ \{ \;  U =_{}^? cons(x, U_1^{}) , \; x =_{}^? g(y, z),
        \; U_1^{} =_{}^? db(V_1^{}, y), \; V =_{}^? cons(y, V_1^{})   \; \}}
        {\eq ~ \uplus ~ \{ \;  U =_{}^? cons(x, U_1^{}), 
                         \; U =_{}^? db(V, z)  \; \}} \]

    \item[(DB5) {\em Flip  $db$ to $bc$ conditionally:} ]
      \[\infer[\quad \mathrm{if} ~V >_c^{+} U, \, ~and ~~V \ngtr_{db}^* \, U]
         { \eq ~ \cup ~ \{ V =_{}^? bc(U, x) \} }
       { \eq ~ \uplus ~ \{ U =_{}^? db(V, x) \} } \]

\end{description}

\medskip
Rules (DB3.a), (DB3.b), (DB4) and (DB5) have the lowest priority: they are 
to be applied in the ``laziest'' fashion. The rule (DB3.b) (``{\em 
Push $bc$ and $db$ below $cons$\dots if $\mathbf{nonnil}$}'') is justified 
by the conditional left-cancellativity of $db$ (cf. Lemma F, {\em Appendix-2}). 
Rule (DB5) is actually a `narrowing' step,  justified by the fact that 
$db$ `is a left-inverse' for $bc$. 

For the completeness of the procedure, we shall also need a few more list 
inference rules which are ``don't-know'' nondeterministic; 
namely, the rules (DB6.a)--(DB8) below:  

\medskip
\begin{description}[font=\normalfont]
    \item[(DB6.a) {\em Guess a Nil-solution-Branch for $db$  
                                 at a $db/db$-peak \/}:]
        \[ \infer
        {\eq ~ \cup ~ \{ U =_{}^? nil, \; V =_{}^? nil, \; W =_{}^? nil \}}
        {\eq ~ \uplus ~ \{ U =_{}^? db(V, x), \; U =_{}^? db(W, y) \}} \]

    \item[(DB6.b) {\em Guess a Nil-solution-Branch for $bc$ and $db$
                                 at a $bc/db$-peak \/}:]
        \[ \infer
        {\eq ~ \cup ~ \{ U =_{}^? nil, \; V =_{}^? nil, \; W =_{}^? nil \}}
        {\eq ~ \uplus ~ \{ U =_{}^? bc(V, x), \; U =_{}^? db(W, y) \}} \]

    \item[(DB7.a) {\em Guess a Narrowing step for $db$ 
                                 at a $db/db$-peak \/}:]      
        \[\infer[\quad \mathrm{if} ~V \ngtr_{db}^* \, U]
         { \eq ~ \cup ~ \{ V =_{}^? bc(U, x), \;  U =_{}^? db(W, y\}\} }
         { \eq ~ \uplus ~ \{ U =_{}^? db(V, x), \; U =_{}^? db(W, y\} } \]

    \item[(DB7.b) {\em Guess a Narrowing step for $db$
                                   at a $bc/db$-peak \/}:]

        \[\infer[\quad \mathrm{if} ~W \ngtr_{db}^* \, U]
         { \eq ~ \cup ~ \{ U =_{}^? bc(V, x), \;  W =_{}^? bc(V, y\}\} }
         { \eq ~ \uplus ~ \{ U =_{}^? bc(V, x), \; U =_{}^? db(W, y\} } \]

    \item[(DB8) {\em Standard Unification on $db$}:]
        \[ \infer
        {\eq ~ \cup ~ \{ U =_{}^? db(W, y), \; V =_{}^? W, \; x =_{}^? y \}}
        {\eq ~ \uplus ~ \{ U =_{}^? db(V, x), \; U =_{}^? db(W, y) \}} \]

\end{description}

\medskip
We denote by $\Inf''_l$ the inference system that extends $\Inf_l'$ 
with the  list-inference rules (DB1)--(DB8), given above. It is 
important to note that the Occur-Check Violation rule (L6) is henceforth to 
be applied to $\dbc$-unification problems in standard form, under the 
partial relation $\succ_l$ \emph{as has been refined above}.

\medskip{}
\begin{prop}~\label{list-inf-dbc-1}
Let $\p$ be any $\dbc$-unification problem, given in standard form.
The inference system $\Inf''_l$  terminates on $\p$ in polynomially 
many steps.
\end{prop}
\proof This is an extension of Proposition~\ref{list-unifiable}, to the 
inference system $\Inf''_l$. The proof of that earlier proposition 
can be carried over practically verbatim: we only have to show that the 
new inferences that might introduce fresh variables, namely the three rules 
(DB3.a), (DB3.b) and (DB4), cannot lead to a non-terminating chain of 
inferences. To ensure this, a first observation is that the relation 
$\sim_{\beta}$, which was used in the proof of Proposition~\ref{list-unifiable},  
{\em has to be refined now} so as to take {\em also} into account the 
relation $\sim_{db}$, the symmetric closure of $>_{db}$, as follows:
\begin{itemize}
\item[-] If $U \sim_{db}^* V$ then $U \mathop{\sim}_{\beta} V$.  
\item[-] Let $U\, >_{cons} \,U'$ and $V\, >_{cons} \,V'$; then 
$U  \mathop{\sim}_\beta  V$ implies $U'  \mathop{\sim}_\beta  V'$. 
\end{itemize}\medskip

\noindent A second observation is that these three rules which might introduce fresh 
variables remove a $\sim_{db}$-edge at some node $U$, and introduce a 
new $\sim_{db}$-edge at a node $U'$ such that $U >_{cons} U'$; but the number of 
${\mathop{\sim}}_{\beta}^{}$-equivalence classes remains the same, by the 
same argument as developed in the proof of Proposition~\ref{list-unifiable}.  
The other details of that earlier proof carry over verbatim.  \qed

\medskip{}
Given any $\dbc$-unification problem $\p$ in standard form, let $\A''$ denote 
the inference procedure based on the rules of $\Inf''_l$, given above 
for its list-equations; we augment the procedure $\A''$ with any given complete 
procedure for solving the residual set of element-equations in the problem, 
when the list-inference rules of $\Inf''_l$ are no longer applicable. 
We have then the following result:  

\medskip
\begin{prop}~\label{Completeness-2}
The procedure $\A''$ is sound and complete for solving $\dbc$-unification 
problems given in standard form. 
\end{prop} 
\proof The proof uses the same lines of reasoning as for 
Proposition~\ref{complete}. The procedure $\A''$ is sound, because 
to any solution of a problem derived under any of its inferences, corresponds 
a solution for the initial problem. The completeness of $\A''$ is again
proved, for any given problem, by induction on the maximum number of 
inference steps needed for the termination of the procedure $\A''$ on the 
problem; and using case analysis  when necessary, based on the ``don't-know'' 
inference rules (DB6.a)--(DB8) above, for such an analysis. We leave out 
the details, which are straightforward.   \qed 

\medskip{}
\begin{prop}~\label{list-inf-dbc-2}
Let $\p$ be a $\dbc$-unification problem in standard form, to which
none of the inferences of $\Inf''_l$ is applicable. Then its subset of 
list-equations with non-nil variables on the left-hand side
is in $d$-solved form.
\end{prop}
\proof This extends Proposition~\ref{d-solved} to the inference system
$\Inf''_l$. Note that we just need to show the following: From any
given node $[U]$ on any given connected component $\Gamma$ of the
Propagation graph $G_l$, there is an unambiguous, cycle-free, directed path
to a well-determined end-node on $\Gamma$. Now, given that any  directed arc on
$G_l$ is labeled with either $>_{cons}$, or $>_{bc}$, or $>_{db}$,
there can be at most one outgoing arc from $[U]$: otherwise one of the
inferences (DB2)--(DB8) would have been applicable; there can be no
directed $\succ_l$-cycle either at $[U]$, otherwise the Occur-Check violation 
rule would have been applicable. Thus, the proof of that earlier proposition
carries over, essentially verbatim.  \qed

\medskip{}
\begin{prop}~\label{DBD-in-NP}
Unification modulo the theory $\dbc$ is  NP-complete.
\end{prop}
\proof Given any $\dbc$-unification problem $\p$, computing a standard form 
can be done in polynomial time (wrt the number of variables of $\p$); the 
same holds also for constructing the propagation graph for the standard form. 
Applying then the inference rules of $\Inf''_l$ till termination,  
on this standard form, takes only polynomially many steps, by 
Proposition~\ref{list-inf-dbc-1}. In case of non-failure, extracting the 
set of element-equations from the resulting problem can obviously be done 
in polynomial time. 

To show that solving $\p$ is in NP, it suffices therefore to show
that the set of its element-equations can be solved, modulo the theory 
defined by the single equation $g(h(x, y), y) = x$, in nondeterministic 
polynomial time. 
But this is a {\em collapsing\/} convergent system, and  the unification 
problem for such theories is known to be decidable and 
finitary~\cite{Hullot,millen96}. 
In particular, a decision procedure can be built by using basic normalized 
narrowing, e.g., as given in~\cite{BaaderSnyd-01}; cf.\ also~\cite{clynch2002}. 
We outline, briefly, such a procedure: 

\medskip
\emph{Procedure for Solving $\E(\p)$}: Note that every equation in $\E(\p)$ 
 is either  a $g$-equation, i.e., an equation of the form $u \Eq g(x, v)$;   
 or an $h$-equation, of the form $u \Eq h(x, y)$. 
 
\begin{enumerate}[label=\arabic*.]
\item IF the set of element-equations is in d-solved form, 
     then return that set; \\ ELSE if the set contains $g$-equations, then 
     go to Step 2; ELSE go to Step 3.  
\item Choose nondeterministically an equation in $\E(\p)$ of 
     the form $u \Eq g(x, v)$; and replace it by the $h$-equation 
     $x \Eq h(u, v)$.
\item If  $\E(\p)$ contains two different $h$-equations with the  
     same lhs variable, apply standard decomposition below $h$ on these 
     two; and suppress one of the two equations. 
\item Apply (element-)Variable Elimination to the resulting set of 
     element-equations, if needed.     
\item Go to Step 1. 
\end{enumerate}

\lft
(Note that Step 2 is just narrowing.) It is easy to check that this 
procedure is in NP on the size of $\E(\p)$.  
 
It remains to show that solving a general $\dbc$-unification problem is 
NP-hard.  This follows from our Proposition~\ref{NPhard} below, where we 
actually make a more precise statement.  \qed 

\medskip{}
\begin{prop}~\label{NPhard}
Unifiability modulo $g(h(x, y), y) = x$ is NP-complete.
\end{prop}

\proof (cf. also~\cite{SivaHaiChrisDranRusi-09}.) We need only to prove the 
NP lower bound; we do that by reduction from the {\em Monotone 1-in-3 SAT\/} 
problem, formulated as follows:
\begin{itemize}
\item[] Given a propositional formula in CNF {\em without negation} 
such that every clause has exactly 3 literals
(variables), check for its satisfiability under the condition
that \emph{exactly\/} one literal in each clause should evaluate to true.
\end{itemize}
This problem is known to be NP-complete~\cite{Schaefer}.

Now consider the following problem of unification modulo 
$g(h(x, y), y) = x$, involving 3 element-variables $x_1, x_2, x_3$:
\disp{$g(h(g(h(g(h(a,b),x_1),b),x_2),b),x_3) =^{?} g(h(a,b),c)$}
where $a, b, c$ are ground constants. 

Since $g(h(x, y), y) \flr x$ is a  convergent rewrite system, the unifiability 
problem is equivalent to finding an instance of the equation under an 
irreducible substitution such that both sides can be reduced to the same 
term. But the right-hand side term $g(h(a,b),c)$ is irreducible modulo  
$g(h(x, y), y) \flr x$; so we need to eliminate two $g$ symbols from 
the left-hand side term $g(h(g(h(g(h(a,b),x_1),b),x_2),b),x_3)$.
The only way to do that is by assigning $b$ to two of the variables, and then 
reduce using the rule $g(h(x, y), y) \rightarrow  x$.
We easily check that we obtain the following possible results: 
$g(h(a,b),x_1)$, $g(h(a,b),x_2)$, $g(h(a,b),x_3)$.
If we assign the third `left-out'  variable  -- let us call it $x$ -- to  
$b$, the term obtained $g(h(a,b),b)$ would reduce to $a$, which is 
irreducible and different from $g(h(a,b),c)$.
If we assign this left-out variable to some irreducible term $t$ different 
from $b$ and $c$, then $g(h(a,b),t)$ would be irreducible, again different 
from $g(h(a,b),c)$. Hence, the only way to reduce both sides of the given 
problem to become equal, is to assign $c$ to the left-out variable. 
In other words: solving this problem amounts to assigning the 
term $c$ to exactly one of the three variables $x_1, x_2, x_3$, and 
assigning $b$ to the other two.

Now let us consider a (finite) set of clauses, each with three 
positive literals. To each clause $L_1 \vee L_2 \vee L_3$ in this set, we 
associate $3$ element-variables $x_1, x_2, x_3$, and the element-equation
$g(h(g(h(g(h(a,b), x_1),b), x_2),b), x_3) =^{?} g(h(a,b),c)$ on these
variables. From the discussion above, the system of derived equations
has a solution modulo $g(h(x, y), y) = x$ if and only if the set of 
clauses is 1-in-3 satisfiable.  \qed

\medskip
\begin{rem}\label{R:4} (i) It can be shown that $\dbc$-unification is finitary, 
along the same lines of reasoning as for the proof of
Proposition~\ref{finitariness}.

(ii) The inference rules (DB5), (DB7.a) and (DB7.b) of $\Inf''_l$ -- which 
are justified by the last equation of $\dbc$ -- play the role of reducing  
unification modulo $\dbc$, in fine, to unification modulo $\bc$. 
\end{rem}

\begin{exa}  (i) The following problem:
$U \Eq db(V, x), \, V \Eq cons(y, W),  \, W \Eq bc(U, z)$
is unsatisfiable. Our procedure exits with failure:  we have an
Occur-Check Violation:  $U \, >_{db} \, V \, >_{cons} \, W \, >_{bc} \, U$.  

\lft
(ii) The following problem $\p$ is in standard form:
\disp{$U \Eq db(V, y), \, U \Eq cons(x, U_1), \, V \Eq cons(y, V_1)$}
We have a $cons/db$-peak at  $[U]$ on the graph of $\p$, and the
only ``don't-care'' rule applicable is the Splitting rule (DB4); we can use
the equation $V \Eq cons(y, V_1)$ for that splitting. After  cancellation on
$cons$ and a variable elimination step, the problem derived is:
\disp{$U \Eq cons(x, U_1), \, x \Eq g(y, y), \, U_1 \Eq db(V_1, y), \,
             V \Eq cons(y, V_1)$}
which is in d-solved form, and gives a solution.    \qed
\end{exa}

\begin{exa} 
(i) The following problem: $U =^? db(V, y), ~~V =^? db(U, z)$ is in 
standard form, but is not in a $d$-solved form. Rule (DB1.c) is applicable, 
and gives the ``nil'' solution to $U$ and $V$, with $y, z$ arbitrary.

(ii) The following problem $\p$ is in standard form: 
$U =^? bc(V, x), ~~ V =^? db(U, y)$, but not in a d-solved form; the only 
applicable inference rule is (DB5) ({\em Flip $db$ to $bc$} conditionally), 
and the problem becomes:
\disp{$U =^? bc(V, x), \hsp U =^? bc(V, y)$}

This is a $\bc$-unification problem which is L-reduced, but not in a d-solved 
form. None of the list-variables $U, V$ is in {\bf nonnil}; so, an obvious easy
solution is $U := nil, \, V := nil$, the element-variables $x, y$  being
arbitrary; this corresponds to applying rule (L8).
We could also nondeterministically apply the rule (L10) 
({\em Standard unification on $bc$}); to deduce then the most general solution 
solution, namely: $U := bc(V, x), x := y$.   \qed
\end{exa} 

\begin{exa}  The following problem $\p$ is in standard (but not 
in a d-solved) form: 
\disp{$U =^? bc(V, x), \, V =^? db(W, y), \, W =^? db(T, z), \, T=^? bc(U, t),
  \, U =^? cons(u, U_1)$}
Observe that $T >_c^{+} W$ but $T \ngtr_{db}^{*} W$, so the rule  (DB5) 
({\em Flip $db$ to $bc$ conditionally}) is applicable to the equation on $W$;
and that gives: 
\disp{$U =^? bc(V, x), \, V =^? db(W, y), \, T =^? bc(W, z), \, T=^? bc(U, t),
  \, U =^? cons(u, U_1)$}
The problem now presents a $bc/bc$-peak at $T$ which is in $\mathbf{nonnil}$, 
so rule (L4.b) can be applied, by writing $W =^? cons(w, W_1)$; this, followed
by Cancellation on $cons$, and a Standard unification step on $h$, leads us 
to deduce: $w =^? u, \, t =^? z, \, W_1 =^? U_1$, and subsequently $W =^? U$; 
the problem is thus transformed (after some Variable Elimination steps) into:  
\disp{$U =^? bc(V, x), \, V =^? db(U, y), \, T =^? bc(U, z), \, 
                    \, U =^? cons(u, U_1), \, W =^? U, \, t =^? z$}

\lft
The rule  (DB5) ({\em Flip $db$ to $bc$ conditionally}) is again applicable, 
now to the equation on $V$; we thus get:
\disp{$U =^? bc(V, x), \, U =^? bc(V, y), \, T =^? bc(U, z), \, 
                 \, U =^? cons(u, U_1), \,  W =^? U, \,t =^? z$}
The rule (L4.a) ({\em Semi-Cancellation on $bc$ at a $bc/bc$-peak}) is now 
applicable, and we deduce: $y =^? x$; after Variable Elimination, the problem 
transforms to: 
\disp{$U =^? bc(V, x), \, T =^? bc(U, z), \, U =^? cons(u, U_1), 
                            \,  W =^? U, \,   y =^? x, \, t =^? z$}
which presents a $cons/bc$-peak on $U$, so the Splitting rule (L5) is
applicable; we write $V =^? cons(v, V_1)$, and the problem evolves (after 
Variable Elimination) to:
\disp{$U =^? cons(u, U_1), \, V =^? cons(v, V_1), \, U_1 =^? bc(V_1, h(v, x)), 
                  \, T =^? bc(U, z), \, W =^? U$,  \par
               \, $u =^? h(v, x), \, y =^? x, \, t =^? z$} 
The  list-equations, as well as the element-equations, are now in $d$-solved
form; and they do give a solution to the problem we started with (as can be
easily checked). \qed

\end{exa}

\section{Conclusion}

\medskip{}
We first addressed the unification problem modulo a convergent 2-sorted 
rewrite system $\bc$, that models, in particular, the (usual, XOR-based) 
CBC encryption mode of cryptography, by interpreting  suitably the function 
$h$ in~$\bc$. 
A procedure is given for deciding unification modulo $\bc$, which has 
been shown to be sound and complete (and finitary) when $h$ is either 
uninterpreted, or interpreted in such a manner. In the uninterpreted case, 
the procedure is a combination of the inference procedure $\Inf_l^{'}$ 
presented in this paper, with syntactic unification; it turns out to be of 
polynomial complexity, essentially for this reason. 
In the case where $h$ is interpreted as mentioned above, the unification 
procedure is a combination of $\Inf_l^{'}$ with any complete procedure 
for deciding unification modulo the associative-commutative theory for XOR; 
and it turns out to be NP-complete for this reason. 
The second part of the work extends $\bc$ into a theory $\dbc$ that 
models, at an abstract level, a cipher-decipher block chaining scheme.  
Unifiability modulo~$\dbc$ is shown to be decidable by an inference 
procedure, which essentially `reduces' any $\dbc$-unification problem 
in fine into one over $\bc$. Unification modulo $\dbc$ is also 
(finitary and) NP-complete.

\medskip
A point that seems worth mentioning here concerns the binary function 
symbol $cons$ in $\dbc$. We have implicitly assumed that in practical 
situations (such as in Example 2 above) the two arguments of 
$cons$ are  `accessible'; this can be made more explicit by adding 
two `projection' equations to $\dbc$, using $car$ and $cdr$ on $cons$, to
get the following set of $8$ equations: 
\begin{eqnarray}
car(cons(x, Y))  & = & x \\
cdr(cons(x, Y))  & = & Y \\  
bc(nil, \, z) & = & nil \\
bc(cons(x, Y), \, z) & = & cons(h(x,z), \; bc(Y, \, h(x,z)))\\
g(h(x,y), \, y) & = & x \\
db(nil, \, z) & = & nil \\ 
db(cons(x, Y), \, z) & = & cons(g(x, z), \; db(Y, \, x)) \\
db(bc(X, y), \, y) & = & X 
\end{eqnarray}
with $car$ typed as $\tau_l \flr \tau_e$, and $cdr$ as $\tau_l \flr \tau_l$. 
All these equations can be oriented left-to-right under a suitable
simplification ordering, and the resulting rewrite system remains convergent. 
It is not difficult to check that, even after the addition of these two 
projection rules, unification problems -- with some very minor 
restrictions on the form of equations involving $car$ and $cdr$ --  
can still be assumed in a standard form, and solved by the inference 
procedure $\Inf''_l$ given above.   
In other words, the results of Section~\ref{DBC} remain valid for this  
enlarged 2-sorted convergent rewrite system -- that we shall again refer 
to as $\dbc$, since no confusion seems likely.   

\medskip
The rewrite system $\dbc$ thus enlarged can actually been shown to be 
$\Delta$-strong in the sense of \cite{siva-dran-micha-07}, under a 
suitable precedence based  (lpo- or rpo- like) simplification ordering, 
by taking $\Delta$ to be the subsystem formed of the two  
rules (6.1) and (6.2).  
It would then follow from Proposition 11 of \cite{siva-dran-micha-07}, that 
the so-called `passive deduction' problem, for an intruder, is decidable, 
if the intruder capabilities are modeled by this theory $\dbc$. 
This would yield, to our knowledge, the first purely rewrite/unification based 
approach for analyzing cryptographic protocols employing the CBC encryption 
mode.  The details will be given elsewhere, where we also hope to present  
decision procedures for a couple of other security problems, where an 
intruder eavesdrops or guesses some low-entropy data in the context 
of block ciphers.   

Finally, observe that unification modulo equational theories often serves 
as an auxiliary procedure in several formal protocol analysis tools,
such as Maude-NPA, CL-Atse, \dots, for handling algebraic properties 
of cryptoprimitives. The work we have presented in this paper 
could be of use in these tools, as a first step towards the automation 
of attack detection in cryptographic protocols employing CBC.  



\pagebreak 
\section*{Appendix-1: On the Cancellativity properties of $bc$, $g$ and $db$}

\medskip\lft
{\bf Lemma A}. For all terms $T_1 , T_2 , t$, we have:  
\disp{ $bc(T_1, t) \approx_{\bc} bc(T_2, t)$ \, if and only if \, 
              $T_1 \approx_{\bc} T_2$.}

\proof The proof is by structural induction on the terms, based on the
semi-cancellativity of $h$ and the cancellativity of $cons$. 
If either $T_1$ or $T_2$ is $nil$, then the other has to be $nil$ too, and 
the assertion of the Lemma is trivial.  So suppose that $T_1$ and $T_2$ 
are not $nil$. Then $T_1 = cons(u_1, T'_1)$ and $T_2 = cons(u_2, T_2')$, for 
some terms $u_1, u_2, T'_1, T'_2$. Substituting back into the original 
equation and applying the second axiom of $\bc$, we deduce that: 
\disp{ $cons(\, h(u_1, t), bc(T'_1, h(u_1, t)) \,)  \approx_{\bc} 
          cons(\, h(u_2, t), bc(T'_2, h(u_2, t)) \,)$ } 

\lft
Since $cons$ is cancellative, we get:
\disp{ $h(u_1, t) \approx_{\bc} h(u_2 , t)$, \,   and  \, 
        $bc(T'_1, h(u_1, t)) \approx_{\bc} bc(T'_2, h(u_2, t))$.}
{From} the  semi-cancellativity of $h$, we then deduce that: 
\disp{ $u_1 \approx_{\bc} u_2 $, \,   and  \, 
        $bc(T'_1, h(u_1, t)) \approx_{\bc} bc(T'_2, h(u_1, t))$.}
Therefore, by structural induction, we deduce that 
$T'_1  \approx_{\bc} T'_2$,  and the result follows.            \qed 

\medskip\lft
{\bf Lemma B}.  For all terms $T, t_1, t_2$, we have:  
\disp{ $bc(T, t_1)  \approx_{\bc} bc(T, t_2)$ \, if and only if \; 
      $T \approx_{\bc} nil$ \, or\, $t_1 \approx_{\bc} t_2$.} 

\proof The proof is by exactly the same reasonings as for proving  
the previous lemma.  \qed

We shall paraphrase these two lemmas together by saying 
that $bc$ is ``conditionally'' semi-cancellative. 

\medskip\lft
{\bf Lemma C}. For all terms $u_1, T_1, u_2, T_2, u_3, u_4$:  
 If $bc(cons(u_1, T_1), u_3) \approx_{\bc} bc(cons(u_2, T_2), u_4)$ \\    
 then \,      
  $h(u_1, u_3) \approx_{\bc} h(u_2, u_4)$ \, and \, $T_1 \approx_{\bc} T_2$.  

\proof By applying the second axiom of $\bc$, we get: 
\disp{$cons(\, h(u_1, u_3), bc(T_1, h(u_1, u_3)) \, ) \approx_{\bc}
                      cons( \, h(u_2 , u_4), bc(T_2, h(u_2, u_4)) \, )$ }
Cancellation on $cons$ gives: 
\disp{$ h(u_1 , u_3) \approx_{\bc} h(u_2 , u_4)$ \, and \, 
            $ bc(T_1, h(u_1 , u_3)) \approx_{\bc} bc(T_2, h(u_2, u_4))$ } 
By Lemma A above, this implies  that  $T_1 \approx_{\bc} T_2$. \qed

In what follows, by $\dbc$ we shall mean the equational theory $\dbc$ of 
Section~\ref{DBC}, and the rewrite system it defines.
   
\medskip
As for the analogs of the above  results for the operator $db$ of $\dbc$, 
we first observe that the function~$g$ is not semi-cancellative -- more
precisely, it is not right-cancellative: indeed, we have  
$g(h(g(t,u), u), u) =_{\dbc}^{} g(t,u)$, although 
$h(g(t,u), u) \not=_{\dbc}^{} t$, in general. But left-cancellativity 
holds for $g$. 
\newpage

\lft
{\bf Lemma D}.~\label{dbc-left-can}
If $g(s, t_1) =_\dbc^{} g(s, t_2)$ then $t_1 =_\dbc^{} t_2$.

\proof We can assume wlog that the terms $s$, $t_1$, and $t_2$ are 
in normal form. If $t_1 \neq_\dbc^{} t_2$, then both $g(s, t_1)$ and 
$g(s, t_2)$ must be redexes, or, in other words, 
$s = h(s', t_1) = h(s', t_2)$ for some~$s'$. Since~$h$
is semi-cancellative this leads to a contradiction. \qed

\medskip\lft
{\bf Corollary E}.~\label{dbc-g-neq}
    If $g(s_1, t_1) =_\dbc g(s_2, t_2)$, 
          and $t_1 \neq_\dbc t_2$, then $s_1 \neq_\dbc s_2$.

\medskip
So, the analog of Lemma A for $db$ does not hold in general. 
However, $db$ is `conditionally' left-cancellative: 

\medskip\lft
{\bf Lemma F}. For all terms $T, x, y$, we have:  
\disp{ $db(T, x) \approx_{\dbc} db(T, y)$ \, if and only if \, 
         $T \approx_{\dbc} nil$ \, or \, $x \approx_{\dbc} y$.}

\proof We just need to prove the ``only if'' assertion. If $T$ is not 
$nil$, then $T = cons(t, T_1)$ for some $t, T_1$. Applying the last 
axiom of $\dbc$, we get: 
\disp{$cons(g(t, x), db(T_1, t)) \approx_{\dbc} cons(g(t, y), db(T_1, t))$.} 
The assertion follows then from the cancellativity of $cons$ and the
left-cancellativity of $g$.   \qed

\medskip{}
\section*{Appendix-2: $db$ as inductive left-inverse for $bc$}

\lft{\bf Lemma G}.\label{db-leftinversefor-bc}
Let $\dbc'$ be the convergent rewrite system formed of the first five 
rules in the system $\dbc$ of Section~\ref{DBC}. 
For any list-term $U$ and element-term $x$ both in  $\dbc'$-normal form, 
we have: $db(bc(U, x), x) = _{\dbc'} U$.

\proof The proof is by structural induction on $U$. The base case when 
$U$ is $nil$ is trivial; so suppose $U = cons(u, U_1)$ for some element-term 
$u$, and list-term $U_1$. Substituting for $U$  and using first the $2$nd 
equational axiom of $\dbc'$, the left-hand side of the assertion becomes: 
\disp{$db(cons( h(u, x), bc(U_1, h(u, x)), x )$.} 
To which we can apply the $5$th equational axiom of $\dbc'$ to get: 
\disp{$cons( g(h(u, x), x), db( bc(U_1, h(u, x)), h(u, x) )$;}
By applying now the $3$rd axiom of $\dbc'$, and the induction hypothesis, 
this reduces  (modulo $\dbc'$) to $cons(u, U_1)$, that is to say $U$. \qed 


\begin{thebibliography}{alpha}

\bibitem{AbadiCortier2004}
 M.~Abadi, V.~Cortier. 
\newblock ``Deciding Knowledge in Security Protocols Under Equational
            Theories''.
\newblock {\em Theoretical Comp. Science}~367(1-2):2--32,~2006.


\bibitem{cbcreport} 
 S.~Anantharaman, C.~Bouchard, P.~Narendran, M.~Rusinowitch.
\newblock ``Unification modulo Chaining''.
\newblock  In {\em Proc.\ of 6th Int. Conference on Language and Automata 
Theory and Applications - LATA 2012}, LNCS 7183, pp. 70--82, 
Springer-Verlag,  2012. 

\bibitem{siva-dran-micha-07}
 S.~Anantharaman, P.~Narendran, M.~Rusinowitch.
\newblock ``Intruders with Caps''.
\newblock In {\em Proc.\  of the Int. Conference RTA'07\/}, LNCS 4533,
pp. 20--35, Springer-Verlag, 2007.

\bibitem{SivaHaiChrisDranRusi-09}
 S.~Anantharaman, H.~Lin, C.~Lynch, P.~Narendran, M.~Rusinowitch.
\newblock ``Unification modulo Homomorphic Encryption''.
\newblock {\em Journal of Automated Reasoning\/}~48(2):135--158~(2012) 

\bibitem{BaaderSnyd-01}
 F.~Baader, W.~Snyder.
\newblock ``Unification Theory''.
\newblock In {\em Handbook of Automated Reasoning\/}, pp. 440--526, 
Elsevier Sc. Publishers B.V., 2001. 

\bibitem{Rogaway1995}
 M.~Bellare, R.~Gu\'{e}rin, P.~Rogaway. 
\newblock ``XOR MACs: New Methods for Message 
             Authentication Using Finite Pseudorandom Function''
\newblock In {\em Proc. \ of the Int. Conference} CRYPT0 '95, 
LNCS 963, pp. 15--28, Springer-Verlag, 1995

\bibitem{Baudet05} 
 M.~Baudet.
\newblock ``Deciding security of protocols against off-line guessing attacks''.
\newblock In {\em Proc.\ of the $12$th ACM Conf. on Computer and Comm. 
Security\/}, CCS'05, pp. 16--25, 2005. 

\bibitem{Comon-T03}
 H.~Comon-Lundh, R.~Treinen.
\newblock ``Easy Intruder Deductions.''
\newblock Verification: Theory and Practice, Essays Dedicated to Zohar 
Manna on the Occasion of His ${64}{\mathrm{th}}$~Birthday
(N.\ Dershowitz, ed.).
\newblock In {\em LNCS~}2772,  pp. 225--242, Springer-Verlag, 2003.

\bibitem{Comon-LICS03}
 H.~Comon-Lundh, V.~Shmatikov.
\newblock ``Intruder Deductions, Constraint Solving and Insecurity 
          Decision in  Presence of Exclusive-Or.''
\newblock In {\em Proc.\ of the Logic In Computer Science Conference,
           LICS'03,\/} pp. 271--280, 2003.

\bibitem{Dersh-JSC}
 N.~Dershowitz.
\newblock ``Termination of Rewriting.''
\newblock {\em Journal of Symbolic Computation\/}~3(1/2): 69--116~(1987).

\bibitem{DolevEvenKarp} D.~Dolev, S.~Even, R.~Karp, 
\newblock ``On the Security of Ping-Pong Protocols''. 
\newblock {\em Information and Control\/}~55:57-68 (1982).

\bibitem{GuoDranWolf-Cade96}
 Q.~Guo, P.~Narendran, D.A.~Wolfram.
\newblock ``Unification and Matching Modulo Nilpotence.''
\newblock In {\em Proc.\ of the 13th Int. Conf. on Automated Deduction,\/} 
  (CADE-13), LNCS~1104, pp. 261--274, Springer, 1996. 

\bibitem{Hullot}
J.-M.~Hullot.
\newblock ``Canonical forms and Unification.''
\newblock In {\em Proc.\ of the 5th Int. Conf. on Automated Deduction,\/}
  (CADE-5), LNCS~87, pp. 318--334, Springer, July~1980.

\bibitem{JoKi}
 J.-P.~Jouannaud, and C.~Kirchner.
\newblock ``Solving Equations in Abstract Algebras: a Rule-Based Survey of
  Unification.''
\newblock In {\em Computational Logic: Essays in Honor of Alan
Robinson,\/}  360--394, MIT Press, Boston, 1991.

\bibitem{KanRev}
 P. C.~Kanellakis, and P. Z.~Revesz.
\newblock ``On the Relationship of Congruence Closure
                    and Unification.''
\newblock {\em J. Symbolic Computation\/}~7: 427-444~(1989).

\bibitem{clynch-liu}
 C.~Lynch, Z.~Liu,
\newblock ``Efficient General Unification for XOR with Homomorphism.''
\newblock In {em Proc.\ of the 23rd Int. Conference on Automated Seduction,\/}
(CADE-23), LNCS~6803, pp. 407--421, Springer-Verlag, 2011.

\bibitem{clynch2002}
 C.~Lynch, B.~Morawska, 
\newblock ``Basic Syntactic Mutation.''
\newblock In {em Proc.\ of the 18th Int. Conference on Automated Deduction,\/}
(CADE-18), LNAI~2392, pp. 471--485, Springer-Verlag, 2002.

\bibitem{millen96}
 J.~Millen, H.-P.~Ko.
\newblock ``Narrowing Terminates for Encryption.''
\newblock In {\em Proc.\ of the Ninth IEEE Computer Security
Foundations Workshop (CSFW)\/}, pp.\ 39--44, 1996.

\bibitem{paterson2011}
 K. G.~Paterson, T.~Ristenpart, T.~Shrimpton. 
\newblock ``Tag Size {\em Does} Matter: Attacks and Proofs for the 
       TLS Record  Protocol''
\newblock In {\em Proc. \ of Int. Conference} ASIACRYPT 2011, LNCS 2073, 
pp. 372--389, Springer-Verlag, 2011. 

\bibitem{Schaefer}
 T. J.~Schaefer.
\newblock ``The complexity of satisfiability problems.''
\newblock In {\em Proc.\ of the 10th Annual ACM Symposium on Theory
of Computing\/}, pp. 216--226, 1978.

\end{thebibliography}
\end{document}